\documentclass[titenlines, reprint, twocolumn, pra,floatfix,groupedaddress,superscriptaddress,aps]{revtex4}
\usepackage{graphicx}
\usepackage{rotating}
\usepackage{array}
\usepackage{multirow}
\usepackage{epstopdf}
\usepackage{amsmath}
\usepackage{amsfonts}
\usepackage{setspace}
\usepackage{braket}
\usepackage{color}

\usepackage{nameref} 

\usepackage{hyperref}

\DeclareMathOperator{\Tr}{Tr}

\begin{document}
\title{
Tensor hypercontraction for self-consistent vertex corrected GW  with  static and dynamic screening;  applications to molecules and solids with superexchange
} 

\author{Pavel Pokhilko}
\affiliation{Department  of  Chemistry,  University  of  Michigan,  Ann  Arbor,  Michigan  48109,  USA}
\author{Chia-Nan Yeh}
\affiliation{Center for Computational Quantum Physics, Flatiron Institute, New York, New York 10010, USA}
\author{Miguel A. Morales}
\affiliation{Center for Computational Quantum Physics, Flatiron Institute, New York, New York 10010, USA}
\author{Dominika Zgid}
\affiliation{Department  of  Chemistry,  University  of  Michigan,  Ann  Arbor,  Michigan  48109,  USA}
\affiliation{Department of Physics, University of Michigan, Ann Arbor, Michigan 48109, USA }

\renewcommand{\baselinestretch}{1.0}
\begin{abstract}
For molecules and solids, we developed efficient MPI-parallel algorithms for evaluating the second-order exchange term with bare, statically screened, and dynamically screened interactions. 
We employ the resulting term in a fully self-consistent manner together with scGW, 
resulting in the following vertex-corrected scGW schemes: scGWSOX, scGWSOSEX, scGW2SOSEX, and scG3W2 theories. 
We show that for the vertex evaluation, the reduction of scaling by tensor hypercontraction (THC) has two limiting execution regimes. 
We used the resulting code to perform the largest (by the number of orbitals) fully self-consistent calculations 
 with the SOX term. 
We demonstrate that our procedure allows for a reliable evaluation of even small energy differences. 
Utilizing a broken-symmetry approach, we explore the influence of the SOX term on the 
effective magnetic exchange couplings. 
We show that the treatment of SOX has a significant impact on the obtained values of the effective exchange constants, 
which we explain through a self-energy dependence on an effective dielectric constant. 
We confirm this explanation by analyzing natural orbitals and local changes in charge transfer quantifying 
superexchange. 
Our analysis explains the structure of weak electron correlation responsible for the modulation of superexchange in both molecules and solids.  
Finally, for solids, we evaluate Neel temperatures utilizing the high-temperature expansion and 
compare the results obtained with experimental measurements.  
\end{abstract}
\maketitle

\section{Introduction}
Green's functions form a traditional language of condensed matter physics enabling the description of complex phenomena, 
such as photoelectron spectrum, conductivity, bulk modulus, and thermodynamic properties at finite temperature, 
as well as numerous one- and two-body properties\cite{Mahan00,Negele:Orland:book:2018,Martin:Interacting_electrons:2016}. 

One of the most common Green's function methods, namely the GW method, was introduced by Hedin~\cite{Hedin65} in 1965. It gave rise to two parent schemes such as fully self-consistent GW (scGW) (usually illustrated as the Hedin pentagram without the vertex term) and scGW+vertex (scGW$\Gamma$) method (illustrated by the full Hedin pentagram).
Both of these methods, scGW and scGW$\Gamma$,  remained for a long time only formal constructs. 
The fully self-consistent zero-temperature GW scheme on the real axis could not be executed fully due to the growing number of poles appearing at every iteration. For a finite temperature on the Matsubara axis, fully self-consistent implementations lacked efficient and compact numerical frequency grids, which made the execution of the scGW scheme possible only for a relatively small number of orbitals.
These serious challenges prevented the fully self-consistent scGW and scGW$\Gamma$ execution for many decades. Instead cheaper and approximate schemes had been commonly employed in computational condensed matter and  materials science: plasmon-pole approximation~\cite{Wu:plasmon_pole:2013}, 
various non-self-consistent GW schemes (G0W0~\cite{Rinke_GW_review}, quasi-particle self-consistency~\cite{QPGW_Schilfgaarde}, eigenvalue self-consistency), and partially self-consistent GW+vertex schemes usually denoted as GW$\Gamma$.
The addition of the vertex correction to GW relies on supplementing the parent GW diagrammatic expansion with additional diagrammatic expressions. Consequently, it is important to decide at which level, self-consistent (scGW) or non-self-consistent (G0W0), the parent GW should be executed. 

From a diagrammatic standpoint, scGW contains bold diagrammatic expansion that contains a partial resummation of an infinite number of diagrams necessary for illustrating important correlations.
Moreover,  important physical laws, such as conservation of energy, momenta, angular momenta, 
gauge invariance of the Luttinger--Ward potential, thermodynamic consistency, 
and current continuity are all guaranteed to be satisfied only 
when the full Green's function self-consistency is achieved~\cite{Baym61,Baym62}.  
Without it, 
Green's function methods may suffer from numerous issues, 
such as a severe dependence on a starting reference, 
an ambiguity in the evaluation of properties, 
or even a violation of the number of particles in the Green's function. 
Yet, despite these diagrammatic advantages, the cost and complication of performing scGW prevented practical schemes of adding vertex corrections. Only very few developments that performed GW$\Gamma$ fully self-consistently have been reported in the past\cite{Kutepov:scGW:CrI3:2021,Kutepov:scG3W2:2022}. 

Consequently, G0W0+$\Gamma$, due to the lower computational cost, remained a much more common approach despite  
theoretical deficiencies. Numerous vertex corrections have been analyzed in the past both for molecular and periodic problems.

In the last 15 years, numerous algorithmic and theoretical advances have resolved technical issues preventing fully self-consistent Green's function methods from being implemented and executed in a timely manner.
Significant advances in the development of compact representations for objects in the Matsubara axis\cite{Kananenka:grids:2016,Kananenka16,Iskakov_Chebychev_2018,dong2020legendrespectral,Yoshimi:IR:2017} have led to the creation of new accurate grids in imaginary time, where now just 100--200 points can be used to represent any fermionic or bosonic quantity with high accuracy yielding highly converged quantities.

At the same time, in the wave-function community, the development  has been focused on 
rigorous and systematically improvable methods that either reduce the calculation cost by introducing  a compact representation of numerical quantities, for example by employing natural orbitals~\cite{Bender:NO:67,Peyerimhoff:NO:74,Peyerimhoff:NO:77, Feller:NOvsHFinCI:92,CDS:98:Rev,Ruedenberg:NOvsHFinCI:02,Sherrill:NOsCI:03,Matsika:FNO:11,Matsika:FNO:13,BARTLETT:FNOvsOVOS:89,Bartlett:FNO:05,Bartlett:FNO:08,Pavel:OSFNO:2019}, or by employing
 tensor decompositions directly reducing the cost of multi-index contractions.  
Examples of such decompositions include 
resolution-of-identity\cite{Whitten:integrals:73,Dunlap:DF:1979,Eichkorn:RIbasis:95-0} and 
Cholesky decompositions of the two-electron integrals\cite{Beebe:Cholesky:77,Koch:Cholesky:2003,Aquilante:Cholesky:2007,Koch:CholMethodSpec:2008,Aquilante:Cholesky:2009,Aquilante:Cholesky2:2009}, 
tensor hypercontraction (THC)\cite{Martinez:THC-MP2:2012,Martinez:LS-THC:2012,Martinez:LS-THC:3:2012,Martinez:THC-CC2:2013,Martinez:THC-CCSD:2014,Martinez:THC-MP2:2015,LinLin:ISDF:2017,Matthews:THC:2020,Lee:THC:2020,Yang:ISDF:THC:2023,Yang:THC-ppRPA:2014,Mardirossian:THC:2018}, 
and canonical polyadic decomposition\cite{Carroll:CP:1970,Hackbush:CP:MP2:2011,Auer:CP:CCSD:2013,Gruneis:CP:CC:2017,Valeev:CP:2021,Valeev:CP:2023}. 
These decompositions vary by their ease of use, 
reduction of cost, and the introduced errors. 
While the resolution-of-identity and Cholesky decompositions 
cannot reduce the overall scaling of the exchange terms and often make their evaluation even more expensive,  
more advanced decompositions, such as THC and canonical polyadic decompositions, 
do provide a reduction in scaling.

Advances in frequency grids coupled with tensor decomposition techniques, such as the resolution of identity (RI), 
have enabled the reduction of both the computational scaling of the direct terms as well as 
the storage requirements for scGW calculations\cite{Iskakov20,Yeh:GPU:GW:2022,Yeh:X2C:GW:2022}. 
Only very recently, THC has been applied to scGW resulting in even lower scaling\cite{Yeh:THC-RPA:2023,Yeh:THC-GW:2024}. 
Additionally, the number of self-consistent iterations in scGW can be reduced due to the introduction of convergence acceleration algorithms~\cite{Pokhilko:algs:2022}. We extended THC for the SOX term with bare interactions for molecules\cite{Pokhilko:THC-GWSOX:2024}. 
In this work, we develop and implement THC algorithms for arbitrary SOX terms with frequency-dependent or frequency-independent interactions in both molecules and solids.

While the influence of the vertex corrections on the band structure was extensively analyzed before within G0W0 schemes, in this paper, we focus on analyzing the influence of the vertex corrections on the energies obtained from the fully self-consistent scGW+vertex schemes. 
To the best of our knowledge such analysis was not attempted before (due to a lack of well-defined energies in the non-self-consistent schemes).

We perform such an energy-based analysis on magnetic systems, where energy differences in broken-symmetry approaches are used to evaluate effective exchange coupling constants.  Such constants are notoriously difficult to evaluate and usually require high-level wave-function methods to get accurate results. Consequently, they are a robust and demanding test of performance for scGW and its vertex corrections.
For this test, we evaluated total energies and energy differences as well as self-energies in scGWSOX, scGWSOSEX, scGW2SOSEX, and scG3W2 theories for several magnetic molecules and solids to gain insight into the effective magnetic couplings using diagrammatic reasoning.

This paper is structured as follows. In Sec.~\ref{sec:theory}, we present an extensive explanation of the concepts used in this paper. This is because the concepts introduced here, such as the self-consistent evaluation of the vertex or evaluation of the exchange coupling constants and energetics were not explored extensively in the GW community. 
To this end in Sec. ~\ref{sec:theory_definitions} and \ref{sec:theory_exchange}, we introduce the necessary theoretical background and explain vertex evaluation techniques within the THC approximation discussed in Sec.~\ref{sec:theory_thc}. We discuss the evaluation of effective exchange couplings in Sec.~\ref{sec:theory_Jcoupling} followed by an explanation of the high-temperature expansion given in Sec.~\ref{sec:theory_high_temp_exp}.
Numerical results are presented in Sec.~\ref{sec:numerical_results} with molecular exchange coupling constants evaluated in Sec.~\ref{sec:results_molecular_j} and Neel temperatures and effective exchange couplings for periodic solids evaluated in Sec.~\ref{sec:results_periodic_j}. Finally, we form conclusions in Sec.~\ref{sec:conclusions}.

\section{Theory}\label{sec:theory}
\subsection{Definitions}\label{sec:theory_definitions}
The imaginary-time one-particle Green's function\cite{Mahan00,Negele:Orland:book:2018,Martin:Interacting_electrons:2016} is defined 
as:
\begin{gather}
G_{\mathbf{pq}} (\tau) = -\frac{1}{Z} \Tr \left[e^{-(\beta-\tau)(\hat{H}-\mu \hat{N})} \mathbf{\hat{p}} e^{-\tau(\hat{H}-\mu \hat{N})} \mathbf{\hat{q}}^\dagger  \right],  \protect\label{eq:G_def}\\
Z = \Tr \left[ e^{-\beta(\hat{H}-\mu \hat{N})} \right], 
\end{gather}
where $\hat{H}$ is the electronic Hamiltonian,
$\hat{N}$ is the particle-number operator, 
$\mu$ is the thermodynamic chemical potential, 
$\mathbf{\hat{p}}$ and $\mathbf{\hat{q}}^\dagger$ are the annihilation and creation operators 
on momentum-dependent spin-orbitals $\mathbf{p}$ and $\mathbf{q}$, respectively, 
$\beta$ is the inverse temperature, 
$\tau$ is the imaginary time at the interval $[0; \beta]$, 
$\Tr$ denotes the trace taken in the Fock space of all possible configurations, and
$Z$ is the grand-canonical partition function. 
For brevity, we combine orbital index, momenta, and spin in a multi-index that we write in bold font: $\mathbf{p}, \mathbf{q}$. 
Due to momenta conservation, the non-zero blocks of $G$ have the same momenta for right and left indices stored as $G^k_{pq,\sigma \sigma}$ and $G^k_{pq,\sigma'\sigma'}$. 
In this work, we consider unrestricted solutions with different $\sigma \sigma$ and $\sigma'\sigma'$ blocks. 
All the expressions for molecular cases can be obtained by letting $k=0$. 

The fully self-consistent methods solve the Dyson equation 
\begin{gather}
G^{-1}(i\omega_n) = G^{-1}_0(i\omega_n) - \Sigma[G](i\omega_n), \protect\label{eq:Dyson}\\
{G}^{-1}_0(i\omega_n) = (i\omega_n +\mu) S - H_0, \protect\label{eq:G0} 
\end{gather}
where $G$ is the full Green's function, $G_0$ is the zeroth order Green's function, 
$\hat{H}_0$ is the zeroth order electronic Hamiltonian matrix expressed in momenta-dependent spin-orbitals, 
$S$ is the overlap matrix of momenta-dependent spin-orbitals, 
$\omega_n$ is the fermionic Matsubara frequency, 
$\Sigma[G]$ is the self-energy functional depending explicitly only on $G$, but not on $G_0$. 

We solve the Dyson equation iteratively and achieve full self-consistency within a small numerical threshold. 
During every iteration, we update the chemical potential to obtain $G$ with the target number of electrons. 

The fully self-consistent methods have a number of important properties 
forming \emph{conserving} approximations\cite{Baym61,Baym62}. 
In particular, energies are conserved and fully self-consistent solutions of the Dyson equation are independent of the starting point. 
These properties are very important for reliable evaluations of energies and energy differences. 
For example, the effective exchange couplings found within non-self-consistent $G_0 W_0$ which in non-conserving approximations were found to be unreliable and heavily dependent on the used DFT references\cite{Chibotaru:BS-G0W0:2020}. In contrast, conserving self-consistent broken-symmetry Green's function approaches gave very systematic effective exchange couplings in both molecules\cite{Pokhilko:local_correlators:2021} and solids\cite{Pokhilko:BS-GW:solids:2022,Pokhilko:Neel_T:2022} comparable with the accurate wave-function theories and experimental measurements.

We evaluate energies through the Galitskii--Migdal expression defined below 
\begin{gather}
\gamma^k_{pq,\sigma\sigma} = -G^k_{qp,\sigma\sigma}(\tau=0^-), \\
E_{\infty}= \frac{1}{N_k} \sum_k \sum_{\sigma} \sum_{pq} \left(h_{pq,\sigma\sigma}+\frac{1}{2}\Sigma^{HF,k}_{pq,\sigma\sigma}[G]\right) \gamma^k_{pq,\sigma\sigma}, \protect\label{eq:E1b}\\
E_{dyn}= \frac{1}{N_k}\frac{1}{\beta}\sum_{\omega_n}\sum_k \sum_{\sigma} \sum_{pq} G^k_{qp}(i\omega_n) \Sigma^{dyn,k}_{pq}[G](i\omega_n), \protect\label{eq:E2b}
\end{gather}
where $E_{\infty}$ and $E_{dyn}$ are the Hartree--Fock and dynamical energies\cite{note:Ecorr} per unit cell evaluated from the correlated one-particle Green's function $G$, 
$\gamma$ is the correlated one-particle density matrix, 
$k$ is a reducible point in the reciprocal space (k-point), 
$\Sigma^{HF}$ is the static (frequency-independent) part of the self-energy functional, 
$\Sigma^{dyn}(i\omega)$ is the dynamic (frequency-dependent) part of the self-energy functional, 
$N_k$ is the volume of the Brillouin zone equal to the total number of reducible k-points used for sampling of the reciprocal space. 
The total energy per unit cell is $E = E_{\infty} + E_{dyn}$. 

\subsection{Second-order exchange with and without screened interactions}\label{sec:theory_exchange}
\begin{figure*}[!h]
  \includegraphics[width=11cm]{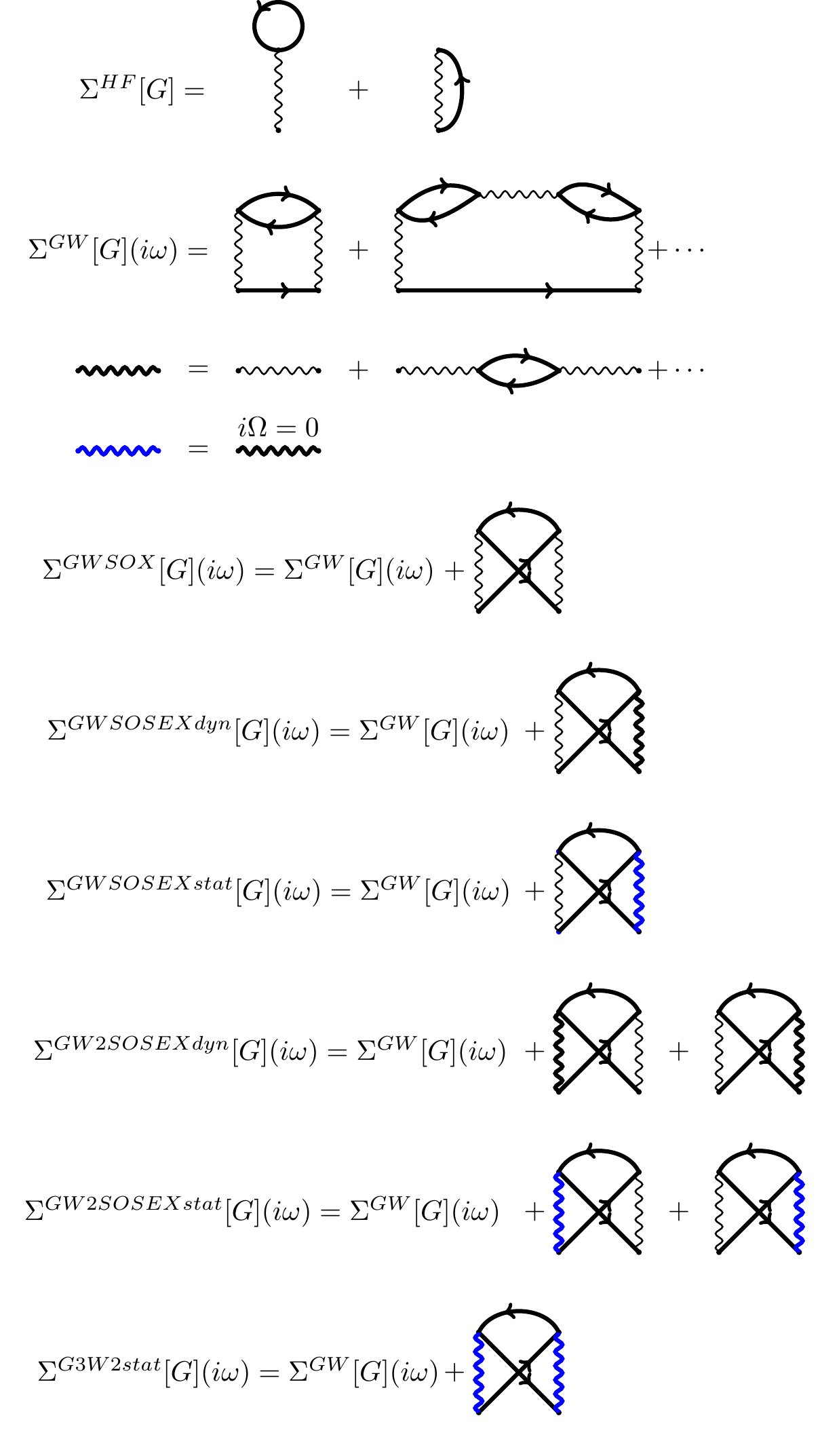}
\centering
\caption{Skeleton perturbative structure of self-energies used in this work. The bold arrow denotes \emph{full} propagator lines $G$ iterated to full self-consistency. 
The thin wavy line is the bare interaction (represented by two-electron integrals). 
The bold black wavy line is the full $W$. 
The bold blue wavy line is the static limit $W(i\Omega=0)$. 
$\Sigma^{HF}$ is the static (Hartee--Fock) self-energy; $\Sigma^{GW}$ is the dynamical part of the GW self-energy. 
The versions of the methods with statically screened interaction $W(i\Omega=0)$ have ``stat'' in their names; 
analogously, the methods with the full $W$ have ``dyn'' in their names.
         \protect\label{fig:diag}
}
\end{figure*}

In practical calculations, the self-energy functional $\Sigma[G]$ in Eq.~\ref{eq:Dyson} is approximated by considering finite diagrammatic expansions such as the one illustrated in Fig.~\ref{fig:diag} based on bold skeleton diagrams. 

The first two terms of this expansion form the static approximation, also known as the Hartree--Fock approximation, which gives the frequency-independent self-energy. 
The next terms give correlated approximations, such as the scGW approximation\cite{Hedin65,G0W0_Pickett84,G0W0_Hybertsen86,GW_Aryasetiawan98,Stan06,Koval14,scGW_Andrey09,GW100,Holm98,QPGW_Schilfgaarde,Kutepov17,Iskakov20,Yeh:X2C:GW:2022,Yeh:GPU:GW:2022}, which has a frequency-dependent part of self-energy and contains an infinite number of direct polarization bubble terms (depicted in Fig~\ref{fig:diag}).  
The polarization bubbles renormalize the bare interaction resulting in 
the screened interaction $W$, denoted through bold thick wavy lines. 
While scGW correctly incorporates the physics of the screened interactions, 
it lacks the second-order Pauli exchange leading to 
two-particle density matrices violating the crossing (permutational) symmetry\cite{Pokhilko:tpdm:2021}.  
The theoretical foundations behind GW are given by Hedin's equations\cite{Hedin65}, 
where an improvement upon the GW approximation is generated by insertions of the vertex functions into the self-energy and polarizability. 
In this paper, we focus on approximating insertions into the self-energy only.

The simplest approximation of the vertex insertion into the self-energy is the second-order exchange (SOX) term, 
resulting in the GWSOX approximation. 
The SOX term completes the second-order exchange diagram by bare interactions. 
The more sophisticated approximations involve screened interactions. 
For example, G3W2 uses two frequency-dependent screened $W$ in the exchange term. This makes G3W2 computationally very challenging and difficult to perform for realistic systems due to involving a convolution in the frequency space. 
Such a theory was executed only by Kutepov using a small number of orbitals\cite{Kutepov:scGW:CrI3:2021,Kutepov:scG3W2:2022} in a fully self-consistent manner, and in Refs.\cite{Rinke:G0W0Gamma0:2021,Foster:SOXstat:2022,Foster:G3W2dyn:2024} in a non-self-consistent manner.
To our knowledge, theories with only one frequency-dependent interaction, SOSEX and 2SOSEX, have been previously considered only within non-self-consistent approximations\cite{Rinke:SOSEX:2015,Foster:SOXstat:2022,Foster:G3W2dyn:2024}. 

In this work, we focus on the assessment of fully self-consistent Green's function theories 
with at most one frequency-dependent $W$ resulting in diagrams shown in Fig.~\ref{fig:diag}. 
In the GWSOSEX approximation, the first interaction is the screened interaction, while the second interaction is the bare one. 
We consider both fully frequency-dependent $W$ (bold black wavy lines) 
and the static approximation of $W$ at $i\Omega=0$ (bold blue wavy lines) commonly used since Hedin's paper\cite{Hedin65}. We distinguish the full and statically screened versions with postfixes ``dyn'' and ``stat'' respectively. 
Ref.~\cite{Foster:G3W2dyn:2024} introduced a non-self-consistent 2SOSEX approximation to G3W2. 
We introduce the fully self-consistent GW2SOSEX approximation with both full $W$ and its statically screened variant. 
Finally, we also consider the fully self-consistent statically screened G3W2 approximation, previously studied only in a non-self-consistent manner.~\cite{Foster:SOXstat:2022,Foster:G3W2stat:2022}

\subsection{THC-SOX and THC-SOSEX: algorithm and implementation details for solids and molecules}\label{sec:theory_thc}
The evaluation of the scGW+vertex terms involves the second-order exchange (SOX) self-energy term with bare or screened interactions. 
When bare interactions are used, the evaluation of this term scales as $O(n_k^3 n_{AO}^5 n_t)$, 
where $n_k$ is the number of k-points in the reciprocal space, $n_{AO}$ is the number of atomic orbitals, $n_t$ is the number of time points from the used frequency grid. 
Such calculations are expensive since the finite-temperature formalism does not distinguish between occupied and virtual orbitals, tensor contractions have to be performed for every frequency point (usually there are 100--200 frequency points), and 
multiple iterations need to be executed until full self-consistency is reached.
Due to these limitations, only a very modest number of orbitals were able to be treated in the previous fully self-consistent vertex calculations in solids and molecules\cite{Phillips14,Rusakov16,Welden16,GF2_Sergei19,Kutepov:scGW:CrI3:2021,Kutepov:scG3W2:2022}.

Recently, we have developed and implemented an MPI-parallel algorithm employing the THC decomposition for the self-consistent evaluation of scGW together with the SOX term with bare interactions. 
This allowed us to apply scGW+SOX to molecules and use it to analyze intermolecular interactions in systems with over 1200 atomic orbitals\cite{Pokhilko:THC-GWSOX:2024}. 
In this paper, we generalize THC-decomposed scGW+SOX to arbitrary static interactions. We also design and implement a new algorithm for evaluating the SOX term with the frequency-dependent screened interactions. 
Both algorithmic advances are tested on solids and molecules allowing examination of larger systems than the ones studied before with fully self-consistent scGW+vertex corrections.

Consistently with the previous papers\cite{Yeh:THC-RPA:2023,Yeh:THC-GW:2024}, we define THC for solids as
\begin{gather*}
(\mathbf{ij}|\mathbf{sr}) = \sum_{PQ}^{N} (X_{iP}^{k_i} )^* X_{jP}^{k_j} U^q_{PQ} (X_{sQ}^{k_s})^* X_{rQ}^{k_r}, \\
q = k_i - k_j + G = k_r - k_s + G^\prime,
\end{gather*}
where $N$ is the number of interpolation vectors, 
$X$ is the collocation matrix, 
the subscript lowercase letters $i,j,s,r$ denote AOs (or other orbitals), 
the subscript capital case letters $P,Q,R$ denote interpolation vectors, 
the superscript $k,q$ denote momenta. Setting the momenta to zero gives a molecular THC decomposition. To perform such decompositions, we used the algorithms presented in Refs.\cite{Yeh:THC-RPA:2023,Yeh:THC-GW:2024} and Ref.\cite{Pokhilko:THC-GWSOX:2024} for solids and molecules, respectively.

\begin{figure}[!h]
  \includegraphics[width=11cm]{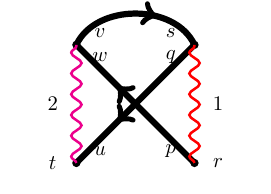}
\centering
\caption{Labeled SOX diagram with arbitrary interactions $1$ and $2$. The bold propagator lines indicate \emph{full} Green's functions and iteration to full self-consistency.
         \protect\label{fig:diag_sox}
}
\end{figure}
Fig.~\ref{fig:diag_sox} shows the most general type of the second-order exchange term that algebraically is expressed as
\begin{gather}
\Sigma_{\mathbf{tr}}^{SOX(U(1),U(2))}(\tau_{tr}) = -\sum_{\mathbf{pqsuvw}}U(1)_{(\mathbf{pr}|\mathbf{qs})}(\tau_{pr,qs}) \nonumber \\ 
U(2)_{(\mathbf{tu}|\mathbf{vw})}(\tau_{tu,vw}) G_{\mathbf{wp}}(\tau_{wp}) G_{\mathbf{uq}} (\tau_{up}) G_{\mathbf{sv}}(\tau_{sv}) \protect\label{eq:sox},
\end{gather}
where $U(1)$ and $U(2)$ are the general time-dependent momentum-dependent interactions,
G is the \emph{full} one-particle Green's function, and the time valuables are $\tau_{xy} = \tau_{x} - \tau_{y}$.

Efficient implementation of Eq.~\ref{eq:sox} depends on the nature of interactions. 
If both $U(1)$ and $U(2)$ are static, there are relations between time points simplifying the time indices 
\begin{gather*}
\tau_t = \tau_u = \tau_w = \tau_v \\
\tau_p = \tau_r = \tau_q = \tau_s \\
\tau_{wp} = \tau_{uq} = \tau \\
\tau_{sv} = -\tau.
\end{gather*}

To reduce computational scaling, 
we generalized our previous molecular THC-SOX algorithm\cite{Pokhilko:THC-GWSOX:2024} for solids as
\begin{gather}
U_{Pqs}^{q_1,k_v} = \sum_Q U(1)_{PQ}^{q_1}(X_{qQ}^{k_v-q_1})^* X_{sQ}^{k_v}\quad O(N^2 n^2_{AO} n_k^2), \protect\label{eq:U_contraction} \\
G_{Rq}^{k} (\tau) = \sum_{uq} X_{uR}^{k} G_{uq,\sigma\sigma}^{k}(\tau) \qquad O(N n_\tau n^2_{AO} n_k),  \\
G_{wP}^{k} (\tau) = \sum_{uq} G_{wp,\sigma\sigma}^{k}(\tau) (X_{pP}^{k})^* \quad O(N n_\tau n^2_{AO} n_k), \\
A_{Pqv}^{q_1,k_v} = \sum_s U_{Pqs}^{q_1,k_v} G_{sv,\sigma\sigma}^{k_v} (-\tau) \qquad O(N n_\tau n^3_{AO} n_k^2), \\
B_{PRv}^{q_1,k_v} = \sum_q A_{Pqv}^{q_1,k_v} G_{Rq}^{q_1-k_v} (\tau) \qquad O(N^2 n_\tau n^2_{AO} n_k^2), \\
C_{RvP}^{q_2,k_w} = \sum_w U(2)_{Rvw}^{q_2,k_w} G_{wP}^{k_w} (\tau)  \qquad O(N^2 n_\tau n^2_{AO} n_k^2), \\
\text{for}\quad q_2,q_1,k_w: D_{PR}^{q_2, k_w-q_2-q_1} \mathrel{+{=}} \nonumber \\
\frac{1}{N_k}\sum_v C_{RvP}^{q_2, k_w} B_{PRv}^{q_1, k_w-q_2} \qquad O(N^2 n_\tau n_{AO} n_k^3), \\
\Sigma_{RP}^{k_t}(\tau) =  -\frac{1}{N_k}D_{PR}^{q_2,k_t-q_2} \qquad O(N^2 n_\tau n_k^2), \\
\Sigma_{tr,\sigma\sigma}^{k_t}(\tau) = \sum_{PR}  (X_{tR}^{k_t})^* \Sigma_{RP}^{k_t}(\tau) X_{rP}^{k_t} \qquad O(N^2 n_\tau n_{AO} n_k). 
\end{gather}
The SOX term contains contractions of only the same-spin quantities, so the spin labels $\sigma\sigma$ are dropped for brevity for most intermediates. 
The reduction of scaling by THC has two regimes determined by $O(N^2 n_\tau n_{AO}^2 n_k^2)$ and $O(N^2 n_\tau n_{AO} n_k^3)$ contractions. This behavior is different from RI-GW, the cost of which is determined only by the contraction with $O(N_{aux}^2 n_\tau n_{AO}^2 n_k^2)$ cost. This difference in regimes has implications for the convergence to a thermodynamic limit by the number of k-points.

Our implementation is MPI-parallel. 
We parallelize over both momenta indices when assembling $U_{Pqs}$ (which is built only at the first iteration and reused in the subsequent iterations), $A$, $B$, and $C$ intermediates. 
Additionally, we parallelize over at least one interpolative vector index. 
We used \textsc{NDA} library\cite{nda} to represent local tensor chunks for every processor.  
We use \textsc{SLATE}\cite{SLATE:2019} for distributed matrix multiplication for $U_{Pqs}$
 and OpenBLAS and MKL BLAS implementations for local matrix multiplications for $A$, $B$, and $C$ intermediates. 
This scheme minimizes communication, most of the computational time is spent on local BLAS operations. 
To save the available memory, we process only a limited number of time points at a time. 

If one of the interactions is time-dependent, the assumptions behind the THC algorithm above are no longer valid 
because of convolutions in the time domain. 
Instead, we implemented the scheme below for such time-dependent interactions, 
which preserves linear scaling with respect to the number of time points for the most expensive contractions
\begin{gather}
GG_{Quv}^{q_1,k_v}(\tau) = G_{uQ,\sigma\sigma}^{k_v-q_1}(\tau) G_{Qv,\sigma\sigma}^{k_v}(-\tau)\quad O(n_t N n^2_{AO} n_k^2), \protect\label{eq:GG} \\
GG_{Quv}^{q_1,k_v}(\Omega_n) = FT(GG_{Quv}^{q_1,k_v}(\tau)) \quad O(n_t^2 N n^2_{AO} n_k^2), \\
A_{Puv}^{q_1,k_v}(\Omega_n) = \sum_Q W_{PQ}^{q_1}(\Omega_n) GG_{Quv}^{q_1,k_v}(\Omega_n) \quad O(n_t N^2 n^2_{AO} n_k^2), \\
A_{Puv}^{q_1,k_v}(\tau) = IFT(A_{Puv}^{q_1,k_v}(\Omega_n)) \qquad O(n_t^2 N n^2_{AO} n_k^2), \\
B_{PRv}^{q_1,k_v}(\tau) = \sum_{u} X_{uR}^{q_1-k_v} A_{Puv}^{q_1,k_v}(\tau) \qquad O(n_t N^2 n^2_{AO} n_k^2), \\
C_{RvP}^{q_2,k_w}(\tau) = \sum_w U(2)_{Rvw}^{q_2,k_w} G_{wP,\sigma\sigma}^{k_w} (\tau)  \qquad O(N^2 n_\tau n^2_{AO} n_k^2), \\
\text{for}\quad q_2,q_1,k_w: D_{PR}^{q_2, k_w-q_2-q_1}(\tau) \mathrel{+{=}} \nonumber \\
\frac{1}{N_k}\sum_v C_{RvP}^{q_2, k_w}(\tau) B_{PRv}^{q_1, k_w-q_2}(\tau) \quad O(N^2 n_\tau n_{AO} n_k^3), \\
\Sigma_{RP}^{k_t}(\tau) =  -\frac{1}{N_k}D_{PR}^{q_2,k_t-q_2}(\tau) \qquad O(N^2 n_\tau n_k^2), \\
\Sigma_{tr,\sigma\sigma}^{k_t}(\tau) = \sum_{PR}  (X_{tR}^{k_t})^* \Sigma_{RP}^{k_t}(\tau) X_{rP}^{k_t} \qquad O(N^2 n_\tau n_{AO} n_k),
\end{gather}
where $FT$ and $IFT$ denote the non-uniform Fourier transform and the non-uniform inverse Fourier transform in the time/frequency domains. 
We use \textsc{sparse-ir} package\cite{Sparse-ir:grid:2023} for time/frequency grids
 and utilize the same time grid for both bosonic and fermionic quantities. 
Since this algorithm assumes storage of all time/frequency points, it demands much more memory. 
To make the dynamical THC-SOSEX/2SOSEX calculations possible for larger systems, 
we move $v$ index to an outside loop making memory demands comparable or even less than in our THC-SOX algorithm with static interactions. 
We again aggressively parallelize over momenta indices, interpolating points, and time/frequency points. 
Consistently with the implementation for static interactions, 
we use \textsc{SLATE}\cite{SLATE:2019} for distributed matrix multiplications.  
However, since the contractions are performed across different axes, 
there is a larger amount of communication involved in the distributed operations. 
In practical examples of molecules and solids below, evaluation of the fully dynamical THC-SOSEX/2SOSEX iterations 
is just 2--3 times more expensive than their static versions.

\subsection{Evaluation of effective exchange couplings}\label{sec:theory_Jcoupling}
Magnetic compounds have a nearly degenerate manifold of electronic states participating in magnetic response which are usually separated from other electronic states. 
This separation allows for downfolding electron correlation into the magnetic subspace of configurations. 
The rigorous theoretical foundation behind this procedure is given by the Bloch formalism\cite{Bloch:1958,Cloizeax:1960,Okubo:1954} that replaces all operators with effective operators preserving any physical observables. Using this formalism, the magnetic Hamiltonians can be explicitly derived.~\cite{Calzado:02,Guihery:2009,Malrieu:2010,Marlieu:MagnetRev:2014,Mayhall:2014:HDVV,Mayhall:1SF:2015,Pokhilko:EffH:2020,Pokhilko:spinchain}
The Bloch formalism relies explicitly on the many-body wave-function amplitudes that are not accessible within the Green's function formalism. 
Instead, to construct the effective Hamiltonians, we follow a simpler strategy based on the broken-symmetry approach formulated first by Noodleman\cite{noodleman:BS:81} and Yamaguchi\cite{Yamaguchi:BS:formulation:1986} within DFT. 
The subsequent publications showed that the broken-symmetry formalism can be utilized within weakly correlated wave-function methods\cite{Yamaguchi:EHF:1988,Yamaguchi:APUMP:1989,Yamaguchi:chapter:instability:1990,Yamaguchi:APCCSD:2012,Stanton:BS-CC:2020}.  
We have shown that broken-symmetry solutions exist in the weakly correlated imaginary-time Green's function methods\cite{Pokhilko:local_correlators:2021}, 
consistent with the generalization of Fukutome's classification of HF solutions\cite{Fukutome:UHF:81} to the correlated solutions of the Dyson equation\cite{Mochena:Fukutome:broken_symmetry:GF}
(but contrary to what was believed from the thermal point of view\cite{Mahan00}). 
We extended the broken-symmetry approach to the series of fully self-consistent Green's function methods and carefully analyzed the implications of the captured electronic structure effects. 
In particular, we have shown that correct capture of electronic screening is paramount to the quantitative description of superexchange in both molecules and solids\cite{Pokhilko:local_correlators:2021,Pokhilko:BS-GW:solids:2022,Pokhilko:Neel_T:2022,Pokhilko:NO:correlators:2023}. In this work, we show that the broken-symmetry solutions of the Dyson equation for molecules and solids also exist for scGWSOSEX, scGW2SOSEX, scG3W2 with static and dynamic screening and analyze the influence of the screened and unscreened SOX term on effective magnetic couplings. 

While within DFT there are numerous spin-decontamination procedures attempting to correct deviations of $\braket{S^2}$ from an ideal one\cite{noodleman:BS:81,Yamaguchi:BS:formulation:1986,Malrieu:spin_pol:BS-DFT:2020,Malrieu:decont:BS-DFT:2020}, 
we have shown that such schemes can be misleading and can introduce additional errors due to the lack of connected contribution to $\braket{S^2}$ 
used in practical DFT calculations\cite{Pokhilko:local_correlators:2021,Pokhilko:BS-GW:solids:2022}.  
Our previous calculations show that spin contamination in the compounds analyzed in this work is negligible. 
Hence, we do not apply such a correction. 
Therefore, our execution of the broken-symmetry approach is reduced to the construction of the \emph{Ising} Hamiltonian from the broken-symmetry (BS) Green's function solutions and solutions with the highest spin projection (HS), 
mapping it to the \emph{Heisenberg} Hamiltonian with the same effective exchange coupling value, 
and reconstructing the magnetic spectrum of the compound. 
For molecules, we use the following definition of the Heisenberg Hamiltonian\cite{note:H_def}
\begin{gather}
H_{eff} = -\sum_{A<B} J_{AB} \vec{S}_A \cdot \vec{S}_B, 
\end{gather}
where $\vec{S}_I$ is the local effective spin operator on the center $I$ and 
the summation is taken over the unique pairs. 
This definition and the evaluation procedure, in particular, imply that in the case of two $1/2$ spins, 
the effective exchange coupling is equal to the singlet--triplet gap
\begin{gather}
J = E(S) - E(T) = 2\left( E(BS) - E(T) \right)
\end{gather}

\begin{figure}[!h]
  \includegraphics[width=7cm]{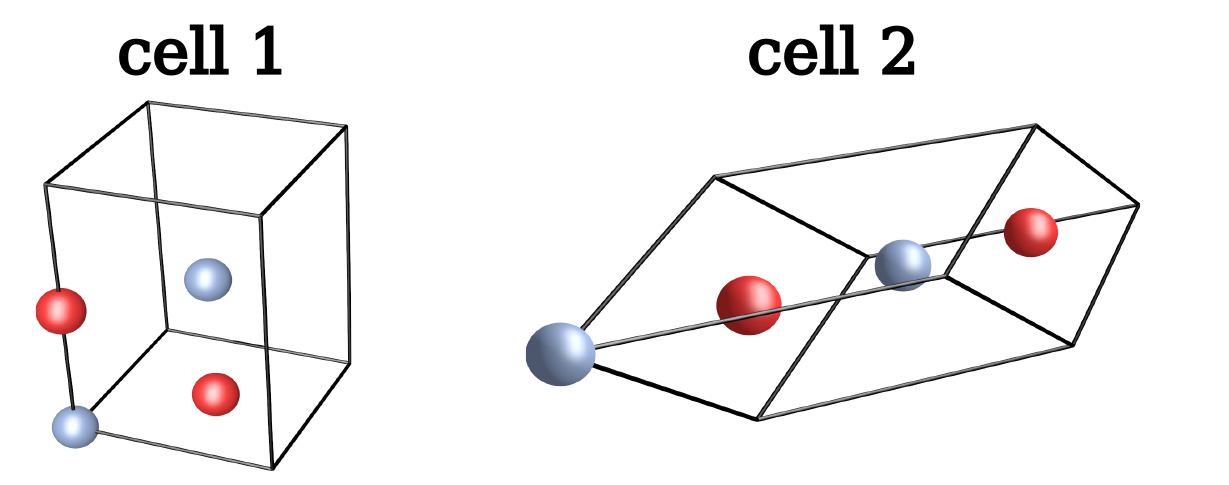}
\centering
\caption{Unit cells used to capture solutions of different types in oxides.
         \protect\label{fig:cell_def}
}
\end{figure}
For solids, another convention is often used with a unit spin normalization. 
Such normalization is convenient for the classical spin limit often used in the approximate treatment of critical temperatures. 
We denote such effective exchange contents with $u$ subscript. 
In particular, the Heisenberg magnetic Hamiltonians for NiO and MnO have non-negligible interactions between the nearest (NN) and next-nearest (NNN) neighbors\cite{Martin:NiO:exchange:2002,Majumdar:NiO:MnO:DFT:J:2011,Pokhilko:BS-GW:solids:2022,Pokhilko:Neel_T:2022}
\begin{gather}
H_{eff} = -J_{1,u} \sum_{\braket{i,j}} \vec{e}_i \vec{e}_j - J_{2,u} \sum_{\braket{\braket{i,j}}}  \vec{e}_i \vec{e}_j,
\protect\label{eq:Ham_defs}
\end{gather}
where again the summation over the unique pairs of NN and NNN is taken and $J_1$ and $J_2$ denote the effective exchange couplings for NN and NNN per formula unit, respectively. 
The molecular convention of the Heisenberg effective Hamiltonian is recovered if the equivalence of energies of 
\emph{Ising} configurations is used $\braket{J_n \vec{S}_i \vec{S}_j} = \braket{J_{n,u} \vec{e}_i \vec{e}_j}$, so 
$J_{n} = J_{n,u}$ for NiO and
$J_{n} = \frac{4}{25} J_{n,u}$ for MnO. 
Although DFT studies often employ large supercells to capture different solutions and to extract the corresponding effective exchange couplings\cite{Majumdar:NiO:MnO:DFT:J:2011}, 
we found a more efficient approach reducing such supercells according to black-and-white symmetries of the solutions\cite{note:magnetic_groups}. 
For NiO and MnO this reduction results in the two cells shown in Fig.~\ref{fig:cell_def}. 
In our previous studies\cite{Pokhilko:BS-GW:solids:2022,Pokhilko:Neel_T:2022} we showed that this description is equivalent to the supercell treatment if one considers energy differences evaluated in the same cell
\begin{gather}
J_{1,u} = -\frac{E(HS1)-E(BS1)}{2\cdot 8}, \\
J_{2,u} = -\frac{E(HS2)-E(BS2)}{2\cdot 6} - J_{1,u},
\protect\label{eq:J_extr}
\end{gather}
where the energies of the solutions are labeled according to the solution type (HS,BS) and the cell where it is found.  
The denominators occur due to the corresponding number of neighbors and signs of $\braket{S_A S_B}$ values. 
The effective exchange couplings are local quantities in insulators; thus, they converge quickly to the thermodynamic limit with an increase of the k-point grid\cite{Pokhilko:BS-GW:solids:2022}.

\subsection{High-temperature expansion and critical temperatures}\label{sec:theory_high_temp_exp}
For molecules with very few magnetic centers, 
temperature-dependent magnetic properties can be evaluated directly 
from the eigenstates of the effective magnetic Hamiltonian as is commonly done in the analysis of EPR spectra\cite{Harriman:EPR:2013}.
Solids often show a more complicated behavior due to the presence of
magnetic phase transitions. The critical temperatures of magnetic phase transitions are difficult to calculate due to several reasons (i) they are sensitive with respect 
to the finite-size effects if estimated from growing finite clusters, (ii)
the direct diagonalization for extended systems quickly becomes impossible because of 
the exponential growth of the dimension of the Heisenberg Hamiltonian matrix, (iii)
approximations based on a classical treatment of spin often result in unrealistic estimates of critical temperatures, 
especially at the low local spin values\cite{Richter:HTE10:code:2014}. 

For extended systems, it is possible to bypass problems with finite-size effects, when employing the high-temperature expansion, introduced by Opechowski\cite{Opechowski:1937}.
This expansion was used within a variety of specific cases of Heisenberg and Ising models\cite{Rushbrooke:Wood:HTE:1955,Wood:Rushbrooke:HTE:1957,Rushbrooke:Wood:Curie:1958,Domb:Sykes:HTE:Curie:1957,Young:triang:1993,Young:kagome:1994}, which illustrate its applicability to critical temperatures. 
The principle of high-temperature expansion is based on the series expansion of thermodynamic quantities, 
such as heat capacity $C$ and magnetic susceptibility $\chi$, around the infinite temperature
\begin{gather}
\chi = \sum_{n=0}^{\infty} c_n \beta^n, \\
C = \sum_{n=0}^{\infty} d_n \beta^n,
\end{gather}
where $c_n$ and $d_n$ are the expansion coefficients and
$\beta$ is the inverse temperature. 
The leading terms in this expansion give rise to the 
Curie--Weiss law of the decay of magnetic susceptibility. 
The high-temperature limit and its associated rate of convergence are not only valid for the magnetic effective Hamiltonians within the canonical ensemble, 
but also for self-energies and Green's functions in 
the grand canonical ensemble.~\cite{Fetter:Walecka:2012,Rusakov14} 
Evaluation of expansion coefficients present in the high-temperature expansion is a combinatorically challenging task, 
which originally was solved only for a particular type of lattices and magnetic interactions at a small order $n$. 
This combinatorial problem was only recently solved up to the order $n=10$ using symbolic algebra, with general assumptions about the number of magnetic interactions 
with decomposition onto spin and lattice components. These general solutions are accessible in the HTE10 code\cite{Richter:HTE8:2011,Richter:HTE10:code:2014}, 
which we use in our work. 
The form of the expansion coefficients does not depend on whether the system is ferromagnetic or antiferromagnetic 
and does not depend on the degree of magnetic frustration. 
Since thermodynamic quantities are not analytical at the critical temperatures, 
the critical temperatures can be estimated from the radius of convergence of the high-temperature expansion. 
To determine the radius of convergence of susceptibility expansion series, 
Rushbrooke and Wood\cite{Rushbrooke:Wood:HTE:1955} used the root test (Cauchy's criterion) and 
Domb and Sykes\cite{Domb:Sykes:HTE:Curie:1957} used the ratio test (d'Alembert's criterion). 

In our earlier work, for cubic rock-salt lattices for a variety of spin values, we obtained the analytical form of the expansion coefficients up to 
$n=10$ converged to the thermodynamic limit 
 and investigated different ways of estimating the radius of convergence.~\cite{Pokhilko:Neel_T:2022}
The expansion coefficients $c_n$ are the homogeneous polynomials on $J_i$ of degree $n-1$, 
which results in the following property of the coefficients
\begin{gather}
c_n(J_1, J_2) = c_n(J_1/J_2, 1) \cdot J_2^{n-1}, \protect\label{eq:homogen_cn}
\end{gather}
where $c_n(J_1, J_2)$ is the coefficient evaluated for the lattice with interaction strengths $J_1$ and $J_2$, 
$c_n(J_1/J_2, 1)$ is the coefficient evaluated for the same lattice but with the interaction strengths $J_1/J_2$ and $1$. 
This equality can be seen as a necessity to obtain the same results regardless of the units used. 
The ratio test is naturally independent on the units used, 
but the original Rushbrooke--Wood root test is not. 
To correct the latter, we introduced a shifted root estimate\cite{Pokhilko:Neel_T:2022} $g_n$ defined as
\begin{gather}
g_n = | c_n|^{1/(n-1)} \to T_N, n \to \infty, \\
q_n = \left|\frac{c_n}{c_{n-1}}\right| \to T_N, n \to \infty, 
\end{gather}
where $q_n$ is the Domb--Sykes ratio estimate and $T_N$ is the critical temperature 
(the Neel temperature in our case).  
We showed that $c_n(J_1, J_2)$ polynomials have zeros in the region where $J_1$ and $J_2$ have different signs. 
In this regime, $q_n$ estimates contain poles and cannot be reliably used. 
This issue does not arise for $g_n$, which are fully applicable in this regime for NiO.  
If $J_1$ and $J_2$ have the same signs, $q_n$ is more preferable because 
under an assumption of a finite critical exponent, $q_n$ allow a Domb--Sykes extrapolation
\begin{gather}
q_n \sim q_{\infty} + \frac{a}{n}, n \to \infty,
\end{gather}
where $q_{\infty}$ is an extrapolated estimate at an infinite expansion order. 
In practice, such an extrapolation is obtained by linear fitting of the few $q_n$ estimates, 
which we use in the current paper for MnO.

\section{Numerical results}\label{sec:numerical_results}
\subsection{Computational details}
We used Dunning's cc-pVDZ\cite{Dunning:ccpvxz:1989,Dunning:ccpvxz:Al-Ar,Duninng:ccpvxz:Sc-Zn} basis set for all molecular calculations and \emph{gth-dzvp-molopt-sr} basis set\cite{GTHBasis} with \emph{gth-pbe} pseudopotentials\cite{GTHPseudo} for periodic solids, consistently with the previous calculations\cite{Pavel:OSFNO:2019,Pokhilko:local_correlators:2021,Iskakov20,Pokhilko:BS-GW:solids:2022,Pokhilko:Neel_T:2022}. 
All the one-electron integrals and initial guesses (UHF) were prepared using PySCF 2.0.1\cite{PYSCF}. 
For molecular calculations, we used even-tempered auxiliary basis sets with parameter beta=2.0 
created by PySCF generator together with the 3-index integrals. 
For molecular calculations, we used a dense Becke grid\cite{Becke1988} for the subsequent pruning and evaluation of THC quadrature similar to Ref.\cite{Lee:THC:2020}. 
For solids, we used Monkhorst--Pack k-point grids\cite{Monkhorst:Pack:k-grid:1976}, Ewald charge-probe finite-size correction of the HF exchange term\cite{EwaldProbeCharge,CoulombSingular},
and the FFT grid from PySCF 2.0.1 and the in-house Cholesky decomposition of integrals with a tight threshold of $10^{-8}$ for the HF part and THC for the correlated part of self-energy.  

We selected a set of transition-metal molecules and solids extensively investigated before 
with DFT\cite{Truhlar:J:SFTDDFT:2011,Orms:magnets:17,Peralta:DFT:2022,Peralta:J:2012,Martin:NiO:exchange:2002,Majumdar:NiO:MnO:DFT:J:2011} and 
ab initio methods\cite{Orms:magnets:17,Pavel:OSFNO:2019,Morokuma:DMRG:biquad_exc:2014,Pokhilko:local_correlators:2021,Stanton:BS-CC:2020,Mayhall:1SF:2015,Zimmerman:Cu:iFCI:2021,Neese:CuAc:2011,Pokhilko:BS-GW:solids:2022,Pokhilko:Neel_T:2022}.
To make comparisons with previous calculations, 
we used the same geometries as previous studies: 
experimental structure of [Fe$_2$OCl$_6$]$^{2-}$ from Ref.\cite{Molins:fe_struct:1998} to compare with Ref.\cite{Morokuma:DMRG:biquad_exc:2014}; 
optimized DFT structures of CUAQAC02 and PATFIA (without ferrocene group) from Ref.\cite{Orms:magnets:17,Pavel:OSFNO:2019};
experimental high-temperature rocksalt structures of NiO and MnO with lattice constants a = 4.1705\AA~\cite{Morosin:NiO:exchange_striction:1971} and 4.4450\AA~\cite{MnO_a}, respectively. 
The broken-symmetry approach within the coupled-cluster method with singles and doubles excitations (CCSD) was executed using PySCF without frozen orbitals to allow a direct comparison with all-electron Green's function calculations. 
The BS-DFT calculations with LDA, PBE\cite{Perdew:96:PBE}, PBE0\cite{Adamo:99:PBE0}, B3LYP\cite{Becke:93:B3LYP} functionals were performed in the same setting as Green's function calculations. 
To make comparisons with previous calculations, the molecular Green's function calculations were performed without frozen core because such a formalism is not yet fully developed within the full self-consistency. 
The Green's functions were executed with intermediate representation grids (prepared with high numerical precision) from \textsc{sparse-ir} package\cite{Sparse-ir:grid:2023} 
with $\Lambda = 10^6$ for molecules and $\Lambda = 10^4$ for solids at inverse temperature $\beta = 1000$~a.u$^{-1}$ with 169 and 104 time points, respectively. 
The Green's function self-consistency is converged with the frequency-dependent CDIIS\cite{Pokhilko:algs:2022} 
and damping within a tight threshold to ensure it does not affect the resulting energy differences. 
Natural orbitals are converted using PySCF tools to a Molden format, 
which we visualized with Gabedit\cite{Gabedit:2011} and rendered with POV-Ray\cite{povray}. 
For the high-temperature expansion, 
we used the previously published polynomials available in Supplementary Information in Ref.\cite{Pokhilko:Neel_T:2022}, 
which were obtained from the $20\times 20\times 20$ lattice (essentially converged to thermodynamic limit), 
effective Hamiltonian from Eq.~\ref{eq:Ham_defs}, custom generation scripts, and HTE10 code\cite{Richter:HTE8:2011,Richter:HTE10:code:2014}.  
We estimate the Neel temperatures from the corresponding HTE convergence radii and compare them with the experimental estimates\cite{Lindgard:NiO:2009,Vernon:NiO:1970,Seehra:NiO:1984,Balagurov:NiO:MnO:2016,Goncharenko:MnO:FeO:neutron_dif:2005,Kubo:antiferromagnetism:1955}. 

\subsection{Convergence with respect to THC decomposition}
\begin{table*} [tbh!]
  \caption{Total energies (a.u.) and effective exchange couplings (K) evaluated with scGW for molecular complexes with different numbers of THC interpolation points.
}
\protect\label{tbl:thc_conv}
\begin{tabular}{l|cc|c||cc|c}
\hline
\hline
   & \multicolumn{3}{c}{[Fe$_2$OCl$_6$]$^{2-}$} &  \multicolumn{3}{c}{PATFIA}  \\
\hline
$N/n_{AO}$ &     HS   &  BS          & $J$, K  &     HS        &      BS        &  $J$, K  \\    
5  &  -5359.745782 & -5359.755207  & -238.1  & -4119.687101 &  -4119.687214 & -71.5 \\  
6  &  -5359.429810 & -5359.439311  & -240.0  & -4119.537343 &  -4119.537447 & -65.7 \\
7  &  -5359.343136 & -5359.352511  & -236.9  & -4119.501919 &  -4119.502022 & -64.8 \\
8  &  -5359.321612 & -5359.330980  & -236.6  & -4119.486742 &  -4119.486845 & -65.1 \\
9  &  -5359.316841 & -5359.326220  & -236.9  & -4119.474836 &  -4119.474938 & -64.6 \\
10 &  -5359.313444 & -5359.322800  & -236.3  & -4119.468991 &  -4119.469092 & -64.0 \\
\hline
\hline
\end{tabular}
\end{table*}
\begin{table*} [tbh!]
  \caption{Total energies (a.u.) and effective exchange couplings (K) evaluated with scGW for NiO solid with different number of THC interpolation points.
}
\protect\label{tbl:thc_conv_solid}
\begin{tabular}{l|cccc|cc}
\hline
\hline
   & \multicolumn{6}{c}{NiO, $3\times 3\times 3$}  \\
\hline
$N/n_{AO}$ &    HS1       &     BS1      &       HS2      &       BS2      &  $J_1$, K & $J_2$, K \\   
 5         & -369.241994  & -369.241170 &   -369.275222 &  -369.280671  &  16.3     &  -159.7  \\   
 6         & -369.225728  & -369.224919 &   -369.258136 &  -369.263536  &  16.0     &  -158.1  \\ 
 7         & -369.216543  & -369.215747 &   -369.250533 &  -369.255920  &  15.7     &  -157.5  \\ 
 8         & -369.214322  & -369.213530 &   -369.248820 &  -369.254202  &  15.6     &  -157.3  \\ 
 9         & -369.213836  & -369.213043 &   -369.248417 &  -369.253797  &  15.6     &  -157.2  \\ 
 10        & -369.213593  & -369.212801 &   -369.248174 &  -369.253554  &  15.6     &  -157.2  \\ 
\hline
\hline
\end{tabular}
\end{table*}
We use THC to evaluate only the dynamical part of the self-energy. The static part of the self-energy is evaluated with 
RI and Cholesky decomposition for molecules and solids, respectively.  

Tables~\ref{tbl:thc_conv} and \ref{tbl:thc_conv_solid} show convergence of total energies 
and effective exchange couplings $J$ with respect to the number of interpolating vectors 
within THC-scGW for molecules and solids. 
Molecular total energies of transition-metal complexes converge much slower than the total energies of solids 
and molecular total energies of compounds with lighter elements from Ref.\cite{Pokhilko:THC-GWSOX:2024}. 
This behavior may be caused by a worse description of very compact core orbitals of transition metals 
by interpolating vectors used in the THC formalism. 
Nonetheless, this effect cancels out when energy differences are evaluated. 
The fast rate of convergence of the effective exchange coupling constants is comparable 
with the rate of convergence of intermolecular interaction energies 
in organic compounds from Ref.\cite{Pokhilko:THC-GWSOX:2024}. 
Based on this convergence behavior, 
we selected $N/n_{AO} = 8$ for molecules and $N/n_{AO} = 7$ for solids for all other calculations. 
These truncation thresholds give an acceptable error of less than 1~cm$^{-1}$ in Tables~\ref{tbl:thc_conv} and \ref{tbl:thc_conv_solid}.

\subsection{Total energies and self-energies}
\begin{table*} [tbh!]
  \caption{Total energies and dynamical part of total energies (a.u.) evaluated with different methods for the HS solution of [Fe$_2$OCl$_6$]$^{2-}$.
}
\protect\label{tbl:Etot}
\begin{tabular}{l|cc|cc|cc}
\hline
\hline
   & \multicolumn{2}{c}{[Fe$_2$OCl$_6$]$^{2-}$}  & \multicolumn{2}{c}{PATFIA} & \multicolumn{2}{c}{CUAQAC02}  \\
\hline
Method       &  $E_{tot}$        & $E_{dyn}$   & $E_{tot}$      & $E_{dyn}$  & $E_{tot}$      & $E_{dyn}$  \\ 
\hline
UHF            &   -5357.324071   &  0         & -4115.131067  & 0         & -4339.316968 & 0         \\ 
scGW           &   -5359.321612   & -3.701486  & -4119.486742  & -7.951968 & -4343.853426 & -8.236229 \\ 
scGWSOX        &   -5358.565043   & -2.095955  & -4117.714597  & -4.268809 & -4342.042567 & -4.475037 \\
scGWSOSEXstat  &   -5358.793082   & -2.581151  & -4118.295975  & -5.477630 & -4342.655709 & -5.749365 \\
scGWSOSEXdyn   &   -5358.943451   & -2.939130  & -4118.571942  & -6.189678 & -4342.923217 & -6.449439 \\
scGW2SOSEXstat &   -5359.032674   & -3.115826  & -4118.894749  & -6.765525 & -4343.284990 & -7.092194 \\
scGW2SOSEXdyn  &   -5359.325003   & -3.811885  & -4119.426460  & -8.134579 & -4343.799086 & -8.437748 \\
scG3W2stat     &   -5358.954857   & -2.928866  & -4118.676206  & -6.274883 & -4343.048838 & -6.571393 \\
CCSD           &   -5359.201960   & ---        & -4119.128933  & ---       & -4343.522980 & ---       \\
\hline
\hline
\end{tabular}
\end{table*}
For the HS solution of [Fe$_2$OCl$_6$]$^{2-}$, Table~\ref{tbl:Etot} shows total energies and dynamical parts of total energies from 
Galitskii--Migdal expression\ref{eq:E2b}. 
The total energy of scGW is below the all-electron CCSD. 
This observation is common and consistent with the previous study\cite{Pokhilko:THC-GWSOX:2024}  
because the GW two-particle density matrix violates permutational symmetry\cite{Pokhilko:tpdm:2021} 
imposed by the fermionic nature of second-quantized operators. 
Incorporation of the bare SOX term overcorrects this violation and raises the total energy too much, 
which is again consistent with the observations from Ref.\cite{Pokhilko:THC-GWSOX:2024}.

We will analyze the effects of the inclusion of different versions of vertices by studying the expression~\ref{eq:E2b}, $E_{dyn}$,  which involves directly only the dynamical self-energy and the Green's function, thus being a direct measure of self-energy changes. 
Electronic screening can be interpreted through an effective dielectric constant renormalizing interactions and reducing interaction strength $W = v / \epsilon$. 
scGWSOSEXstat replaces one of the bare interactions in the SOX term with the statically screened interaction, 
which reduces the overcorrection of the exchange term $\Sigma^{SOSEX} \sim \frac{1}{\epsilon} \Sigma^{SOX(v,v)}$ 
and results in a total energy that can be placed between scGW and scGWSOX total energies.

scG3W2stat statically screens both interactions $\Sigma^{G3W2stat} \sim \frac{1}{\epsilon^2} \Sigma^{SOX(v,v)}$ and reduces the total energy further from scGWSOSEXstat.  
The total energy of scGW2SOSEX is below scG3W2, which can be explained from the structure of self-energy. 
The full $W$ has frequency-independent and frequency-dependent components 
$W = v + v\Pi(\Omega) v + \cdots = v + W_d(\Omega)$. 
The G3W2stat self-energy is 
\begin{gather}
\Sigma^{G3W2stat} = \Sigma^{SOX(v,v)} + \Sigma^{SOX(v,W_d(\Omega=0))}  + \nonumber \\ \Sigma^{SOX(W_d(\Omega=0),v)} + \Sigma^{SOX(W_d(\Omega=0),W_d(\Omega=0))} = \nonumber \\
\Sigma^{SOSEX}  + \Sigma^{SOX(W_d(\Omega=0),v)} + \Sigma^{SOX(W_d(\Omega=0),W_d(\Omega=0))} = \nonumber \\
\Sigma^{2SOSEX} + \Sigma^{SOX(W_d(\Omega=0),W_d(\Omega=0))}.
\end{gather}
The analysis in terms of an effective dielectric constant gives a more sophisticated scaling $\Sigma^{2SOSEX} \sim \frac{2-\epsilon}{\epsilon} \Sigma^{SOX(v,v)}$, which can revert the sign of the second-order self-energy if the effective dielectric constant is large enough. 
This argument is valid for both static and dynamic treatments of screening in the SOX term. 
This analysis explains why $E_{dyn}^{scGW2SOSEXdyn} < E_{dyn}^{scGW}$ for [Fe$_2$OCl$_6$]$^{2-}$, PATFIA and CUAQAC02, 
but $E_{dyn}^{scGW2SOSEXstat} > E_{dyn}^{scGW}$ for the same compounds. 
\begin{figure}[!h]
  \includegraphics[width=7cm]{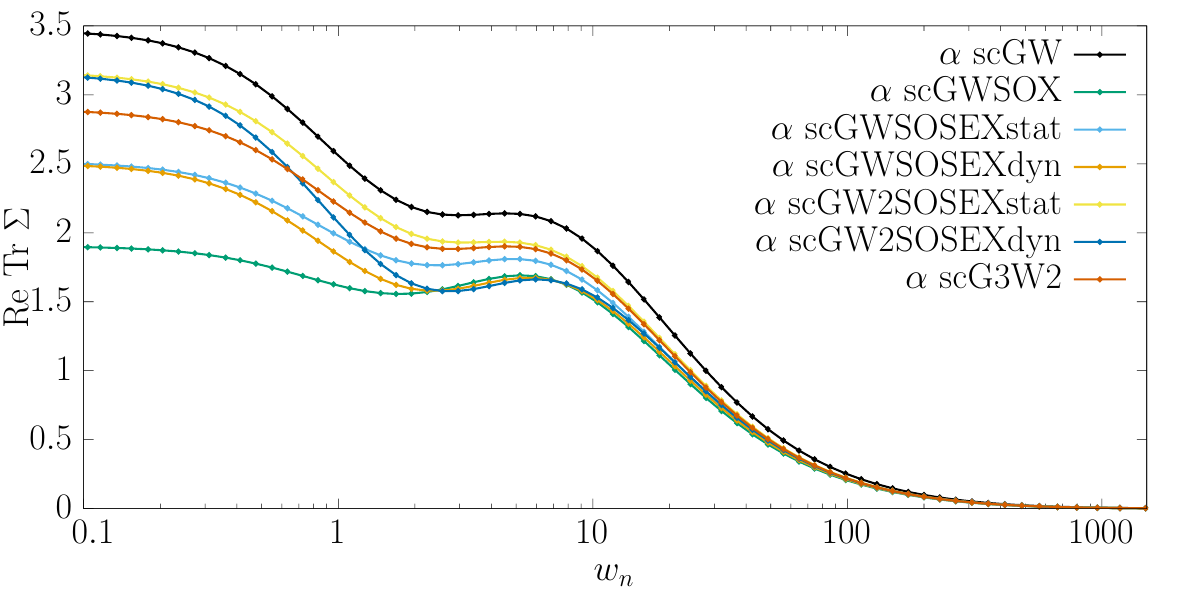}
  \includegraphics[width=7cm]{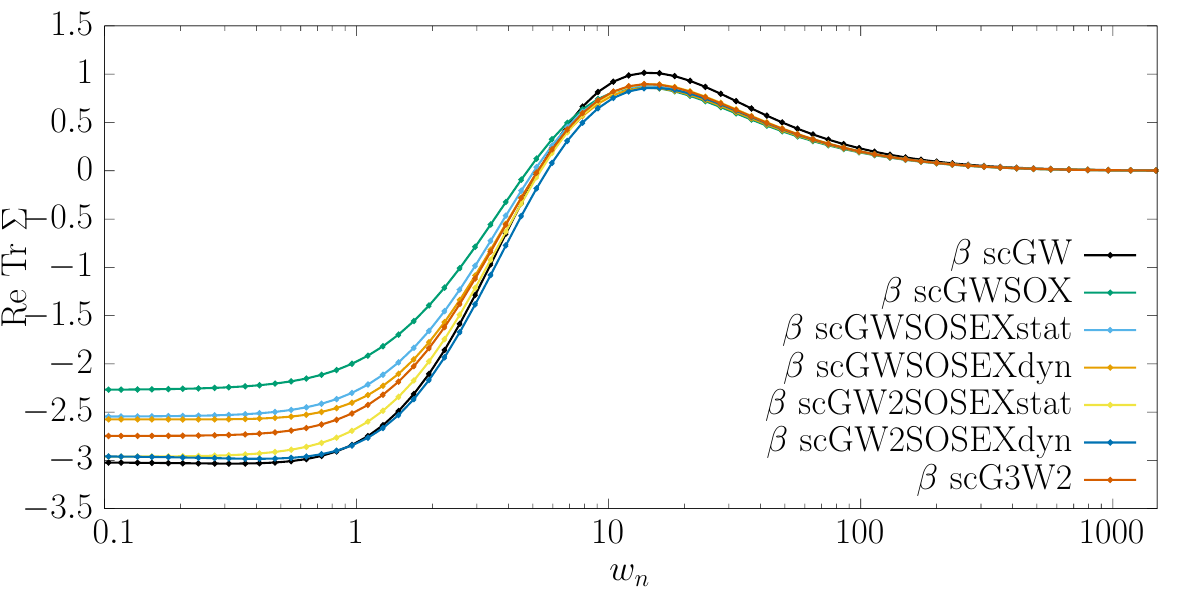}
  \includegraphics[width=7cm]{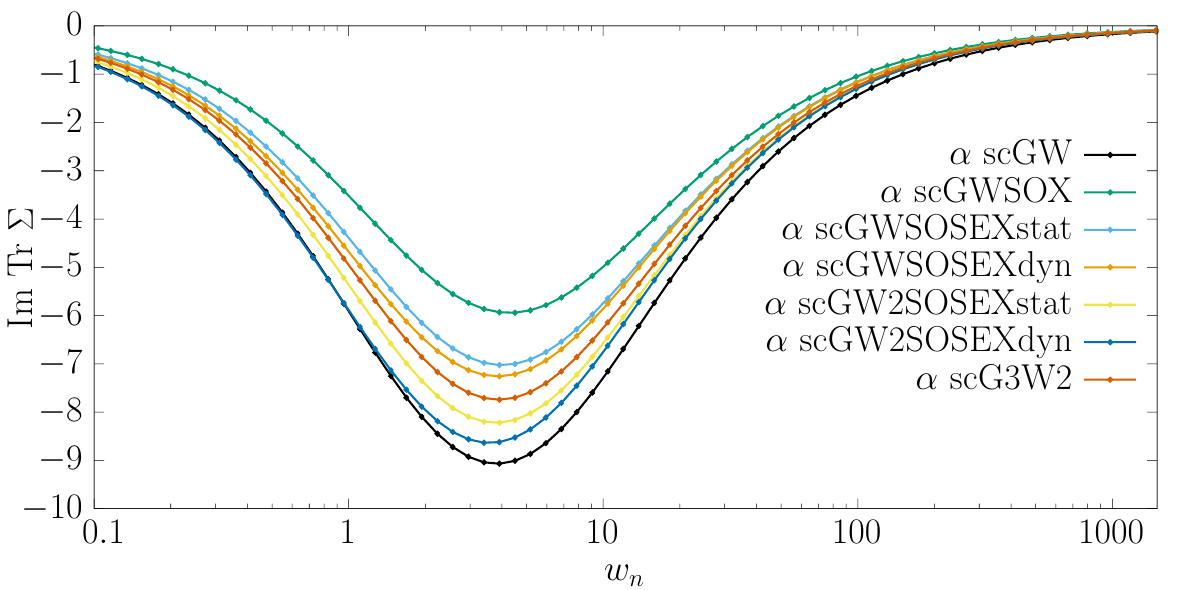}
  \includegraphics[width=7cm]{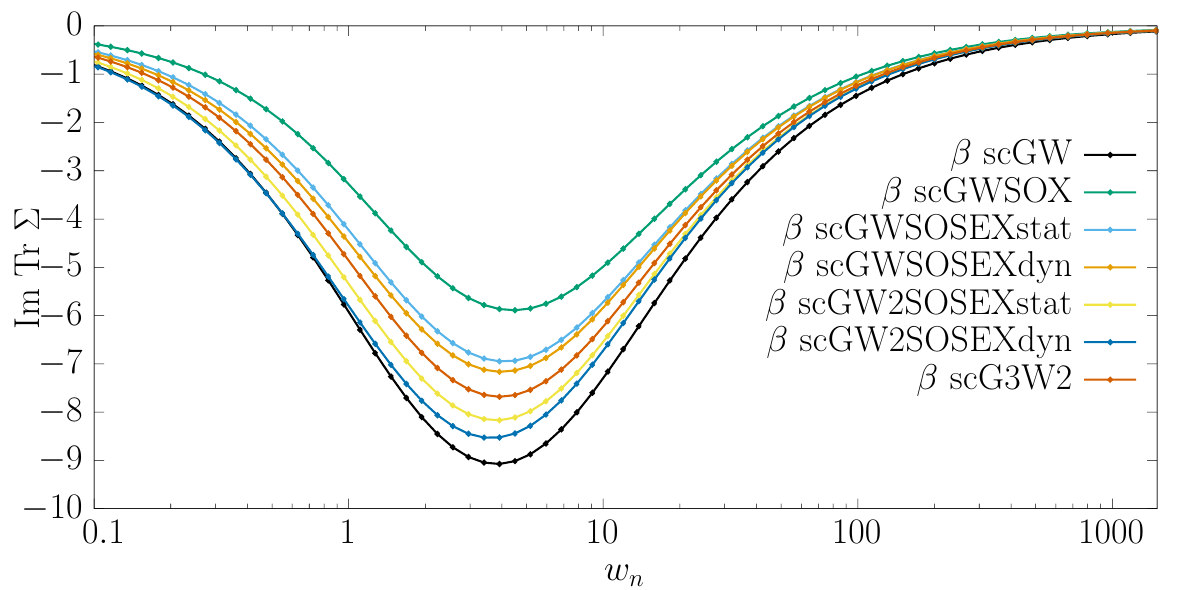}
  \includegraphics[width=7cm]{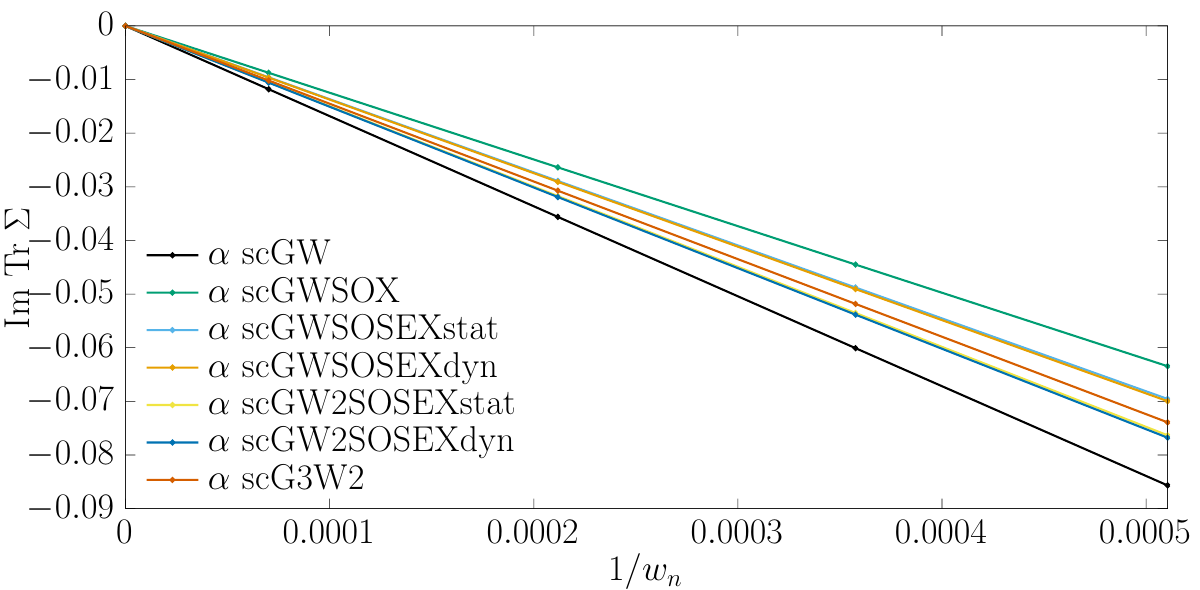}
  \includegraphics[width=7cm]{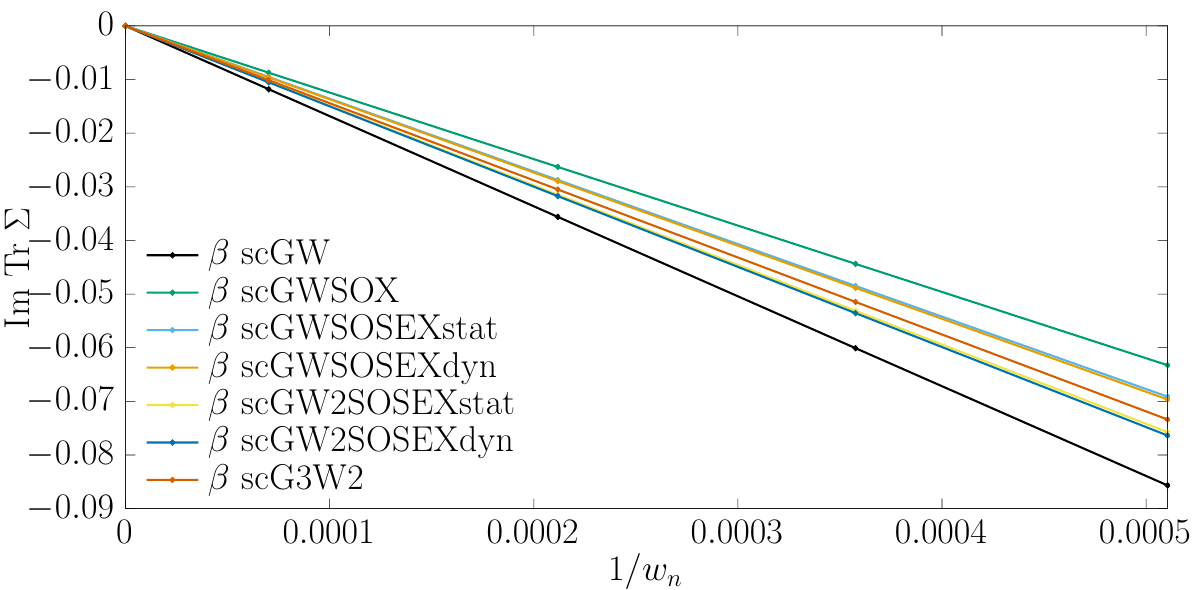}
  \centering
\caption{Traces of real and imaginary components of $\Sigma(i\omega_n)$ of the HS solutions evaluated with different methods for [Fe$_2$OCl$_6$]$^{2-}$.           \protect\label{fig:sigma_w}
}
\end{figure}

A closer inspection of the self-energy gives a trend consistent with $E_{dyn}$. 
Fig.~\ref{fig:sigma_w} shows real and imaginary components of the trace of self-energy in 
Lowdin-orthogonalized orbitals for [Fe$_2$OCl$_6$]$^{2-}$. 
Most of the self-energies with the screened SOX are strictly between GW and GWSOX. 
The statically screened self-energies (GWSOSEX and GW2SOSEX) are close to the 
corresponding dynamically screened self-energies, especially at extreme values of $\omega_n$; 
the only deviation between them occurs in the middle of the frequency range where the real self-energy parts of the dynamically screened methods give a different slope; the imaginary parts are also noticeably different.   
These features can explain why the total energies of the statically and dynamically 
screened methods are significantly different. 
The ordering of slopes of self-energies at the high-frequency limit 
is the same as the ordering of self-energies around the minimum of imaginary components in Fig.~\ref{fig:sigma_w}. 
The imaginary part of scG3W2stat self-energy is between scGW2SOSEX and scGWSOSEX; the same is true for the real part as well near the low values of $\omega_n$. 
The GW2SOSEX self-energies are not strictly bound by scGW self-energies, which is consistent with the inequality between their $E_{dyn}$. 

\subsection{Effective exchange couplings in molecules}\label{sec:results_molecular_j}
\begin{table*} [tbh!]
  \caption{Effective exchange couplings (listen in K) evaluated in non-self-consistent broken-symmetry calculations starting from scGW Green's functions.
}
\protect\label{tbl:J_nonsc}
\begin{tabular}{l|c}
\hline
\hline
 [Fe$_2$OCl$_6$]$^{2-}$  & $J$, K  \\
\hline
BS-GWSOX@scGW               & 4883  \\    
BS-GWSOSEXstat@scGW         & 1749  \\ 
BS-GW2SOSEXstat@scGW        & -2274 \\  
BS-G3W2stat@scGW            & 390   \\ 
\hline
\hline
\end{tabular}
\end{table*}
We compare exchange coupling constants evaluated from non-self-consistent vertex calculations with the ones evaluated from fully self-consistent vertex calculations in Tables~\ref{tbl:J_nonsc} 
and ~\ref{tbl:J_sc_mol}. 
Table~\ref{tbl:J_nonsc} shows effective exchange constants for [Fe$_2$OCl$_6$]$^{2-}$ 
evaluated with the non-self-consistent vertex methods.  In non-self-consistent methods, we used the vertex containing self-energy that was evaluated as a one-shot procedure using the fully self-consistent GW Green's function. The total energy was evaluated from Eq.~\ref{eq:E1b} and \ref{eq:E2b}. We denote such one-shot non-self-consistent calculations through ``@'' indicating the Green's function used for the one-shot self-energy evaluation. 
The resulting non-self-consistent exchange coupling constants values are unphysical and are far from both experimental and theoretical estimates. 
Most of such non-self-consistent calculations fail to predict even what the ground state is. 
The poor quality of non-self-consistent estimates is very comparable with the previous non-self-consistent $G_0 W_0$ estimates from Ref.\cite{Chibotaru:BS-G0W0:2020} which concluded that non-self-consistent Green's function calculations cannot be reliably used even to determine trends in single-molecule magnets. 

\begin{table*} [tbh!]
  \caption{Effective exchange couplings (K) evaluated with fully self-consistent broken-symmetry Green's function calculations and broken-symmetry CCSD for molecules from this work compared with the previous wave-function calculations and estimates extracted from the experimental temperature dependence of magnetic susceptibility and inelastic neutron scattering. 
Due to the used convention for $J$, the listed numbers for PATFIA and CUAQAC02 are identical to the singlet--triplet gaps, $E_{S}-E_{T}$. 
}
\protect\label{tbl:J_sc_mol}
\begin{tabular}{l|p{4.2cm}|c|p{3.8cm}}
\hline
\hline
 $J$, K    & [Fe$_2$OCl$_6$]$^{2-}$  & PATFIA  &  CUAQAC02 \\
\hline
Exp                &  -337$^a$       & -16$^b$ &  -411$^c$,-420$^d$,-411$^e$,\newline-434$^{e}$,-428$^{f}$      \\
BS-UHF             &  -63.7          & -2.1    &  -63.3      \\ 
BS-scGW            &  -236.6         & -65.1   &  -258.3     \\ 
BS-scGWSOX         &  -113.8         & +10.2   &  -136.4     \\ 
BS-scGWSOSEXstat   &  -189.0         & -25.1   &  -204.5     \\ 
BS-scGWSOSEXdyn    &  -194.9         & -21.8   &  -212.4     \\ 
BS-scGW2SOSEXstat  &  -341.3         & -138.7  & -486.2      \\ 
BS-scGW2SOSEXdyn   &  -345.5         & -113.4  & -435.6      \\ 
BS-scG3W2stat      &  -231.5         & -50.9   & -264.9      \\ 
\hline
BS-CCSD            &  -269.3$^g$;-279$^h$ & -74.6$^g$   & -308.4$^g$      \\ 
EOM-SF-CCSD        &  -340.7$^i$     & -122$^j$    & -274$^j$        \\
other              &  $-180$..$-156^k$ CASSCF $-360$..$-326^k$ MRCI+Q $-336$..$-275^k$  DMRG   & $-255^l$  iFCI(n=3)  & $-27.2^m$ CASSCF\newline $-57.3^m$ NEVPT2\newline  $-97.1^m$ DDCI2\newline  $-390.3^m$ DDCI3 \\
\hline
LDA   & -606.2$^g$ & -694.3$^g$ & -1976$^g$ \\
PBE   & -707.4$^g$ & -707.8$^g$ & -1747$^g$ \\
PBE0  & -353.8$^g$ & -98.2$^g$  & -553.3$^g$ \\
B3LYP & -406.7$^g$ & -128.3$^g$ & -678.6$^g$ \\
\hline
\hline
\end{tabular}

$^a$ Ref. \cite{Molins:Fe2OCl6:magnet:2003}

$^b$ Ref. \cite{lopez:PATFIA:05}, should be interpreted with caution

$^c$ Ref. \cite{CUAQAC02:struct}

$^d$ Ref. \cite{Elmali:CuAc:2000}

$^e$ Ref. \cite{CuAc:J:exp:1956}

$^f$ Ref. \cite{Furrer:CuAc:1979}

$^g$ This work

$^h$ BS-CCSD/cc-pVDZ with approximate spin contamination correction from Ref.\cite{Stanton:BS-CC:2020}

$^i$ Estimate with 6-31G* from Ref.\cite{Mayhall:1SF:2015}

$^j$ CD-EOM-SF-CCSD/cc-pVDZ from Ref.~\cite{Pavel:OSFNO:2019}

$^k$ The values extracted in different ways from Ref. \cite{Morokuma:DMRG:biquad_exc:2014}

$^l$ Estimate with 6-31G* from Ref. \cite{Zimmerman:Cu:iFCI:2021}

$^m$ Ref. \cite{Neese:CuAc:2011}
\end{table*}

Exchange coupling constants evaluated from fully self-consistent scGW+vertex schemes (denoted with ``sc'') shown in Table~\ref{tbl:J_sc_mol} are qualitatively different from the non-self-consistently evaluated coupling constants. 
First, they are much closer to the previous theoretical and experimental estimates. 
Second, they elucidate the fundamental physics of the problem.  
This drastic difference from non-self-consistent calculations is not surprising because 
a full self-consistency removes the reference dependence and maintains energy conservation leading to the equivalence of energy evaluation from different expressions.

The strength of the interaction grows substantially from BS-UHF to BS-scGW. 
We have observed this trend before for both molecules\cite{Pokhilko:local_correlators:2021} and solids\cite{Pokhilko:BS-GW:solids:2022,Pokhilko:Neel_T:2022} and 
showed that the origin of this phenomenon is in the captured physics using two-particle local charged correlators and natural orbitals\cite{Pokhilko:local_correlators:2021,Pokhilko:NO:correlators:2023}. 
Two complimentary descriptions can be used to understand superexchange: 
localized orbitals (describing degrees of covalency and charge transfer) 
and delocalized orbitals (describing bonding). 
An effective dielectric constant created by the polarization propagator reduces effective interactions, 
thus, effectively reducing energies of ligand-to-metal charge-transfer contributions, 
which are the cause of superexchange.  
This makes the presence of such contributions much more prevalent in the broken-symmetry solutions causing 
redistribution of the population of the frontier natural orbitals. 
Such a redistribution of population is forbidden by the Pauli principle in the solution with the highest spin projection. 
Hence, differences between both one- and two-particle densities between two solutions increase from BS-UHF to BS-scGW
which, in turn, leads to an increase in the energy differences between the solutions and the resulting magnitude of $J$. 

The delocalized nature of natural orbitals as well as their structure allows one to interpret superexchange 
as a partial formation of multi-center multi-electron bonds. 
Among a variety of different orbital schemes analyzing effective couplings\cite{Kahn:book:1993}, 
natural orbitals\cite{Yamaguchi:magnetic:NO:2000,Malrieu:mag_orbitals:2002,Morokuma:DMRG:biquad_exc:2014,Gagliardi:Cr2muOH:superexchange:2020,Pokhilko:local_correlators:2021,Orms:magnets:17,Kotaru:Fe:SF-DFT:2023} 
are the most informative ones because they represent one-electron density matrix in the most compact way. 
Spin-averaged natural orbitals (SA-NOs, eigenvectors of $\gamma_{\alpha\alpha}+\gamma_{\beta\beta}$) 
are especially useful for quantifying superexchange\cite{Pokhilko:local_correlators:2021,Pokhilko:NO:correlators:2023} 
in UHF as well as correlated wave-function and Green's function methods.

\begin{figure*}[!h]
  \includegraphics[width=14cm]{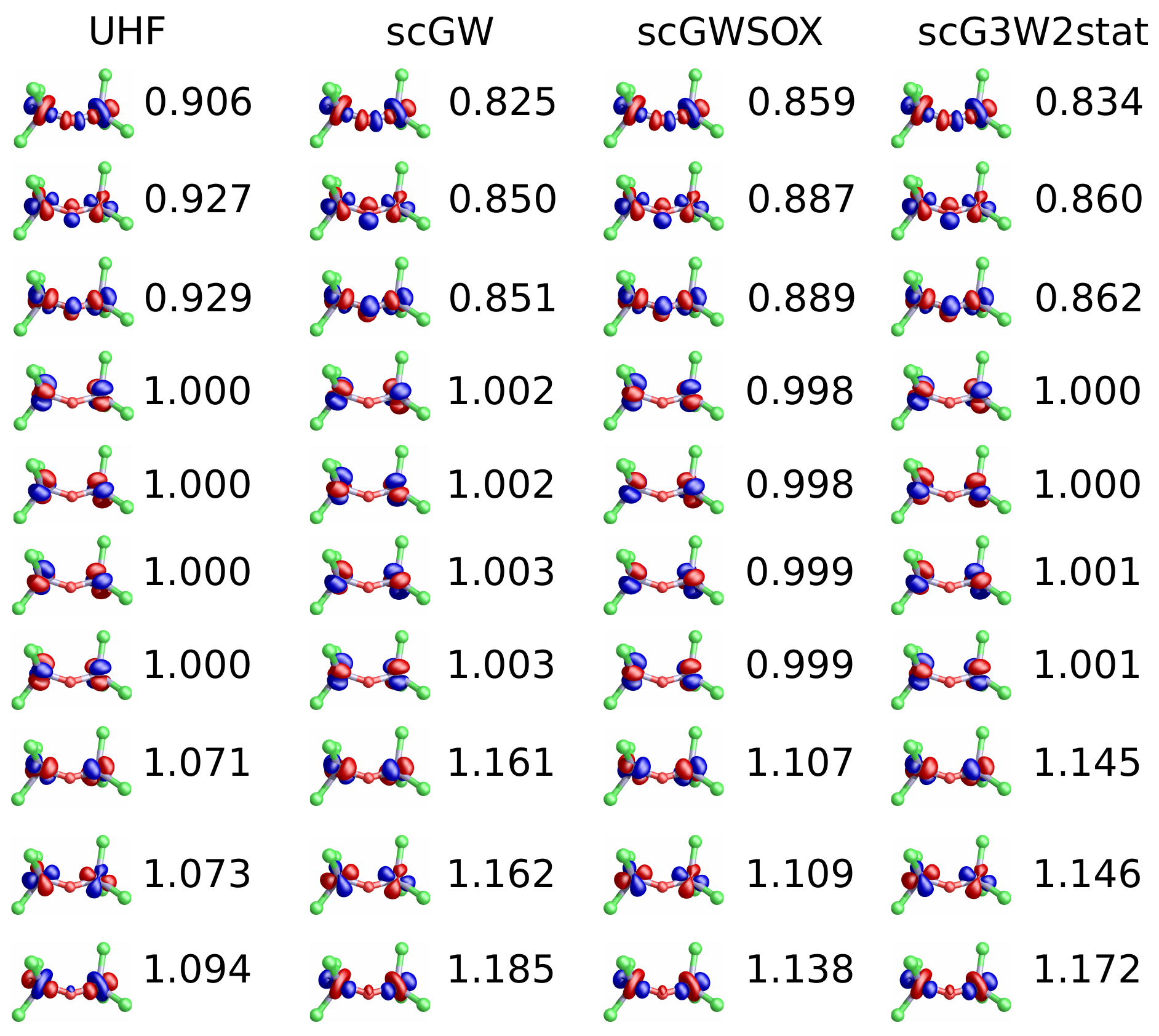} 
\centering
\caption{Spin-averaged frontier natural orbitals found with different methods and their occupancies 
for the broken-symmetry solutions of the Dyson equation for [Fe$_2$OCl$_6$]$^{2-}$. 
Charge transfer is enhanced by the bubble terms and (partly) suppressed by the (screened) SOX term . 
         \protect\label{fig:fe_NO1}
}
\end{figure*}

Fig.~\ref{fig:fe_NO1} shows SA-NOs for [Fe$_2$OCl$_6$]$^{2-}$ obtained in different self-consistent scGW+vertex schemes.
We plot the contributions of $d$-orbitals on Fe and $p$-orbitals on O 
to the SA-NOs with antibonding character as well as to the non-bonding orbitals. 
The occupations of these SA-NOs deviate from 1.   
A partial depopulation of antibonding orbitals (Fig.~\ref{fig:fe_NO1}, the top 3 orbitals) and additional occupation of the corresponding non-bonding orbitals (Fig.~\ref{fig:fe_NO1}, the bottom 3 orbitals) in the broken-symmetry solution happens already in UHF. 
Since the corresponding non-bonding orbitals do not have a contribution from the bridging oxygen, 
this picture corresponds to a partial charge transfer in terms of localized orbitals.
The incorporation of electronic screening makes charge transfer much more prominent 
moving the population from the antibonding orbitals to the corresponding non-bonding ones. 
In addition, the weights of the $p$-orbitals on O become larger, 
corresponding to an enhanced charge-transfer character (Tab.~S1 in SI).  
Solutions with the highest spin projections give the SA-NOs, occupancies which are very close to 1 for all methods. 
Due to the numerical degeneracy of these occupancies, 
the corresponding eigenvectors can be mixed in an arbitrary manner making their visualization ambiguous.  

\begin{figure*}[!h]
  \includegraphics[width=14cm]{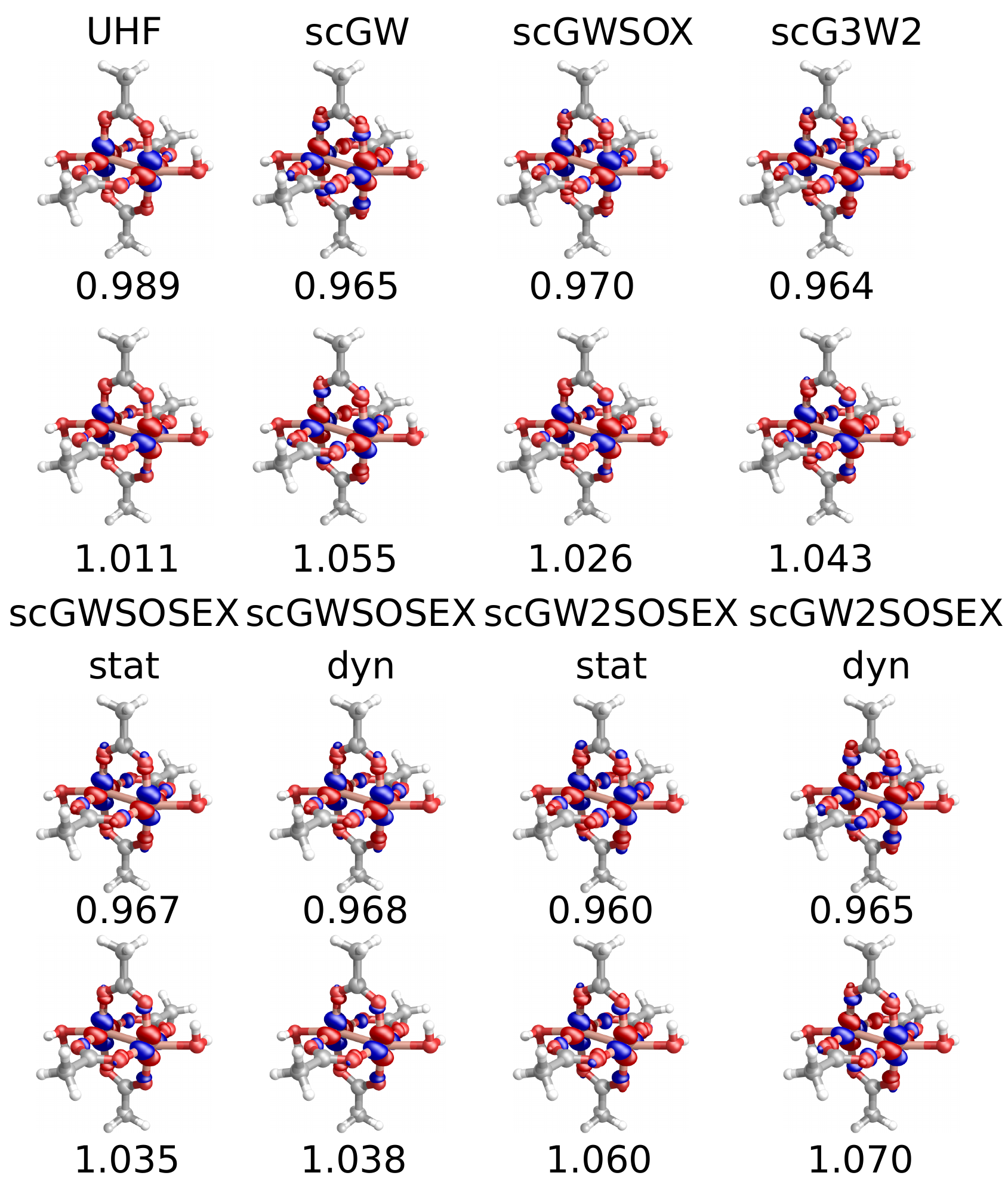}
\centering
\caption{Spin-averaged frontier natural orbitals found with different methods and their occupancies 
for the broken-symmetry solutions of the Dyson equation for CUAQAC02.  
The dynamic and static treatment of screening in SOX is shown. 
         \protect\label{fig:CuAc_NO2}
}
\end{figure*}

The SA-NOs for copper complexes, illustrated in Fig.~\ref{fig:CuAc_NO2}, show similar trends, 
although the absolute values of the losses and gains of the population are smaller than for [Fe$_2$OCl$_6$]$^{2-}$. 

The bare SOX term has a different sign from the direct second-order term. 
Since the bare SOX term is not renormalized, it overestimates the effect of Pauli repulsion 
for all the electrons with the same spin.~\footnote{This becomes clear from the structure of the two-particle density matrix derived by the application of Hellmann--Feynman theorem to the grand potential functional\cite{Pokhilko:tpdm:2021}. }
Since this contribution is subtracted from the one caused by screening, 
SOX effectively reduces the effect of screening, which is also consistent with the change in self-energies shown in Fig~\ref{fig:sigma_w}.  
This observation fully explains why the absolute values of the effective exchange constants evaluated with scGWSOX are much smaller than the ones evaluated with scGW.

Changes in natural orbitals and their occupancies are also consistent with an effective reduction of screening. 
Oxygen contributions to antibonding SA-NOs in Fig.~\ref{fig:fe_NO1} are smaller for scGWSOX than for GW (see also weights of oxygen's p-orbitals in Tab.~S1 in SI). 
The occupancies of antibonding and the corresponding non-bonding orbitals for scGWSOX are closer to 1 than for scGW, 
which also indicates a substantially reduced charge-transfer character. 
The changes in occupations of SA-NOs in copper complexes are also consistent with this picture. 
For PATFIA, scGWSOX even changes the ordering of the solutions predicting a weak ferromagnetic coupling. 
\begin{figure*}[!h]
  \includegraphics[width=14cm]{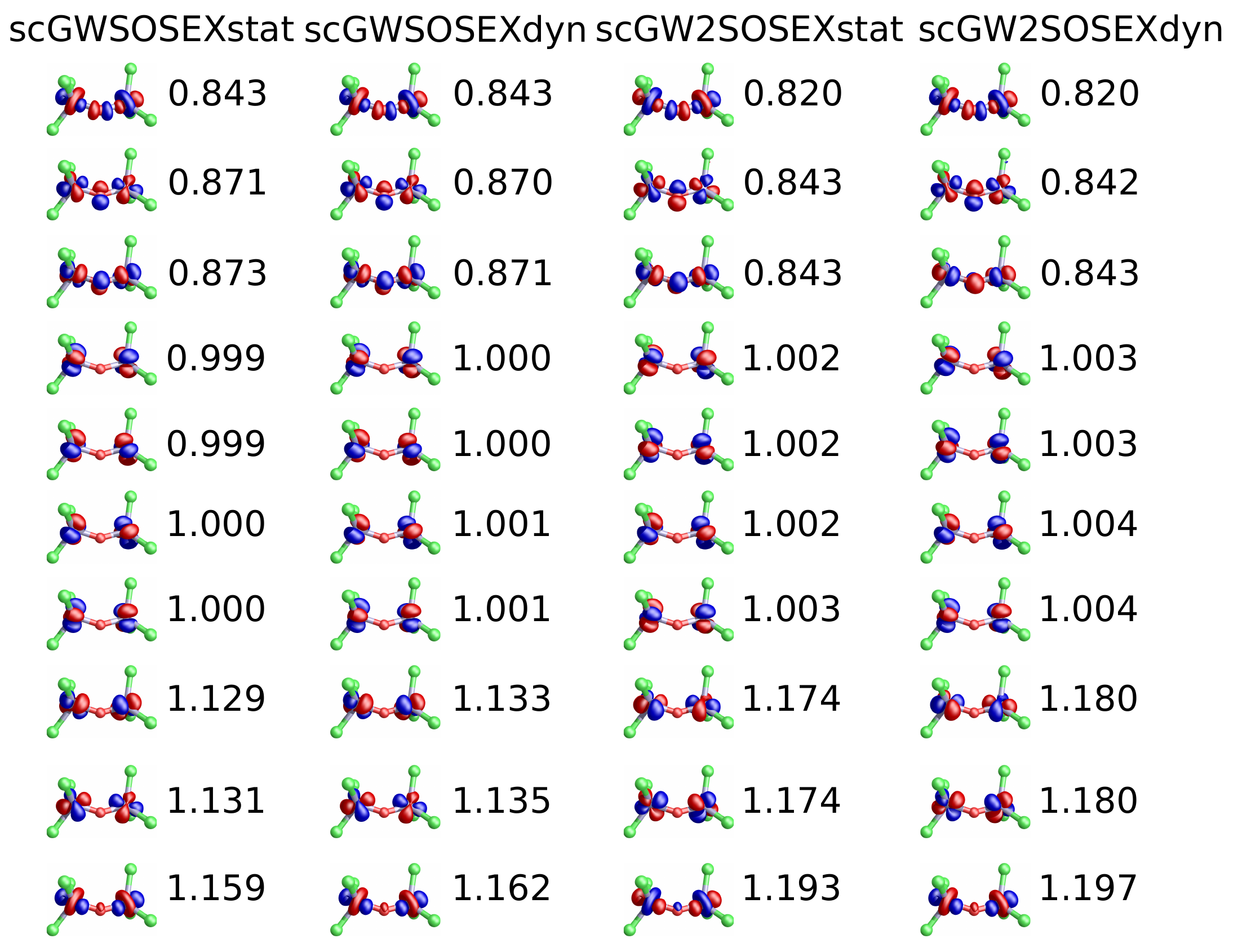}
\centering
\caption{Spin-averaged frontier natural orbitals found with different methods and their occupancies 
for the broken-symmetry solutions of the Dyson equation for [Fe$_2$OCl$_6$]$^{2-}$. 
The dynamic and static treatment of screening in SOX is shown.
         \protect\label{fig:fe_NO2}
}
\end{figure*}
\begin{figure*}[!h]
  \includegraphics[width=14cm]{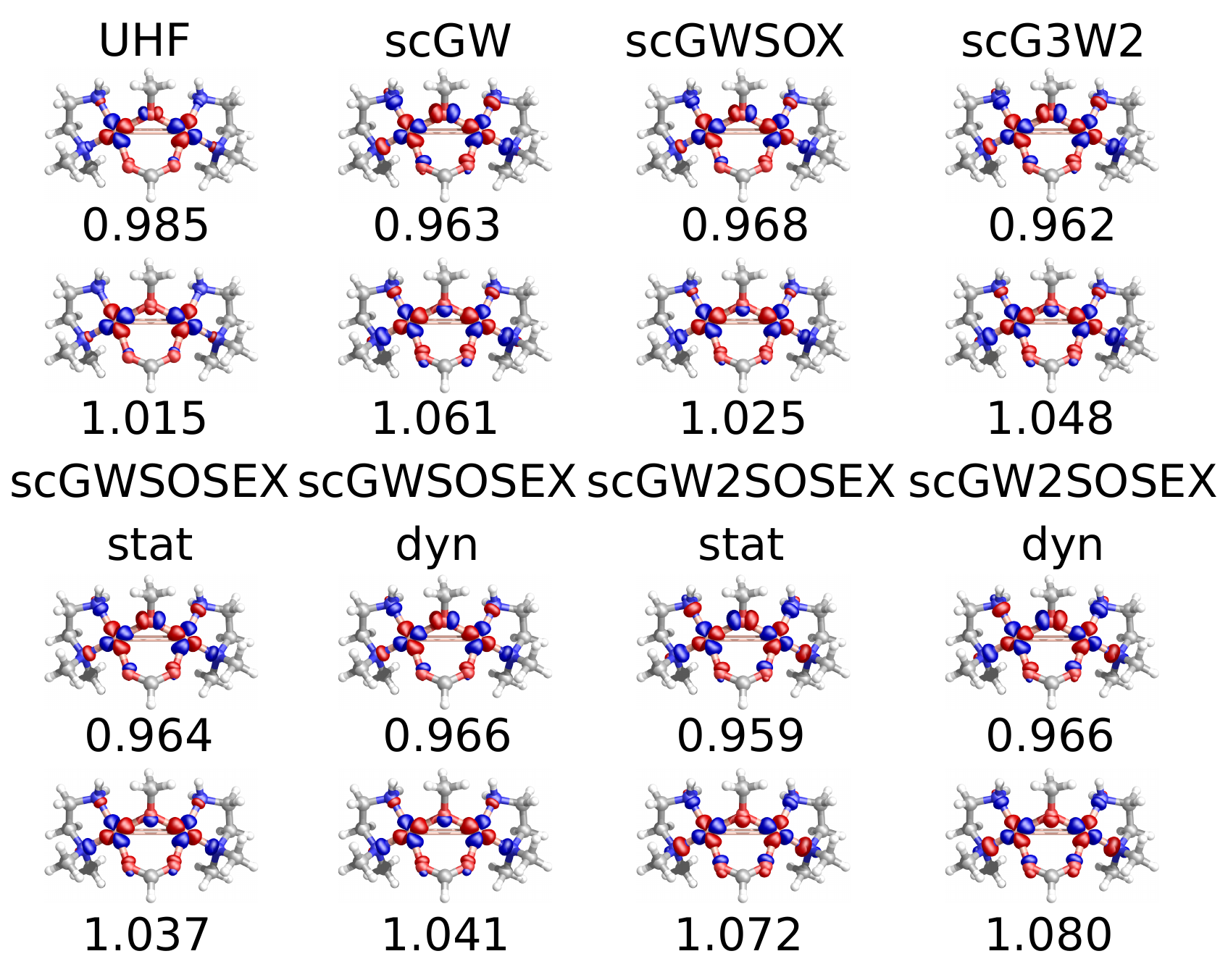}
\centering
\caption{Spin-averaged frontier natural orbitals found with different methods and their occupancies 
for the broken-symmetry solutions of the Dyson equation for PATFIA.  
The dynamic and static treatment of screening in SOX is shown. 
         \protect\label{fig:patfia_NO}
}
\end{figure*}

Replacement of one of the bare interactions in SOX with the screened one in SOSEX reduces 
the overestimation of Pauli repulsion that is caused by using the bare SOX term.
In contrast to the total energies, 
the fully dynamical treatment gives estimates of energy differences 
close to the static approximation indicating that 
the static approximation of $W$ in SOX is a very good approximation in the 
low-energy regime that we target in this work. 

Additionally, SA-NOs evaluated with the dynamical and static treatments of $W$ are very similarly looking as illustrated in Fig~\ref{fig:fe_NO2}. 
The scGWSOSEX estimates of $|J|$ are 
systematically smaller than the scGW ones, 
but the scGW2SOSEX estimates are systematically larger than the scGW ones. 
These trends are consistent with the trends in $E_{dyn}$ as well as 
with our previous explanation regarding the sign and magnitude of the correlated Pauli repulsion. 
Occupancies of frontier SA-NOs are also consistent with these effects. 
For [Fe$_2$OCl$_6$]$^{2-}$, the occupancies of the 3 antibonding SA-NOs and also the 3 corresponding non-bonding SA-NOs fulfill the inequality
\begin{gather}
|1-\sigma_i^{UHF}| < |1-\sigma_i^{scGWSOX}| < |1-\sigma_i^{scGWSOSEX}| < \nonumber \\
< |1-\sigma_i^{scG3W2}|  < |1-\sigma_i^{scGW}| < |1-\sigma_i^{scGW2SOSEX}|, 
\end{gather}
where $\sigma_i$ is the occupancy of the frontier $i$-th SA-NO orbital. 
This sequence of inequalities is not fulfilled fully for the 4 non-bonding orbitals with almost unchanged occupancy. 
For the copper complexes, all except the last one or two inequalities are fulfilled.  
The GW2SOSEX estimates for CUAQAC02 are more sensitive to the treatment of $W$, 
albeit the effect of this approximation is still much smaller than the choice of the method. 

\begin{figure}[!h]
  \includegraphics[width=9cm]{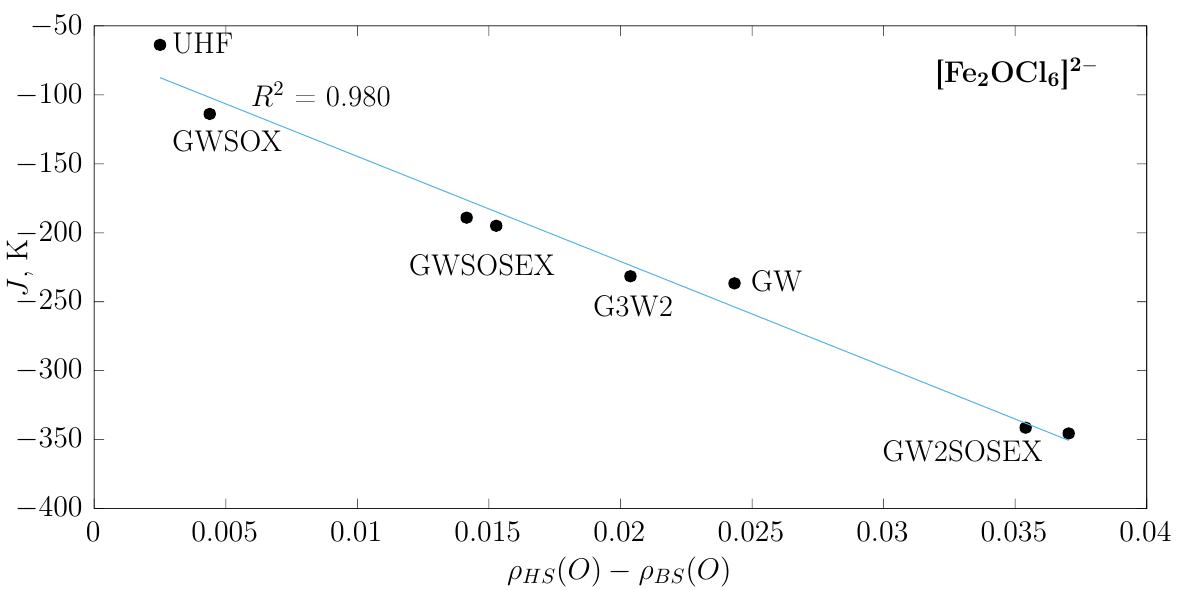}
  \includegraphics[width=9cm]{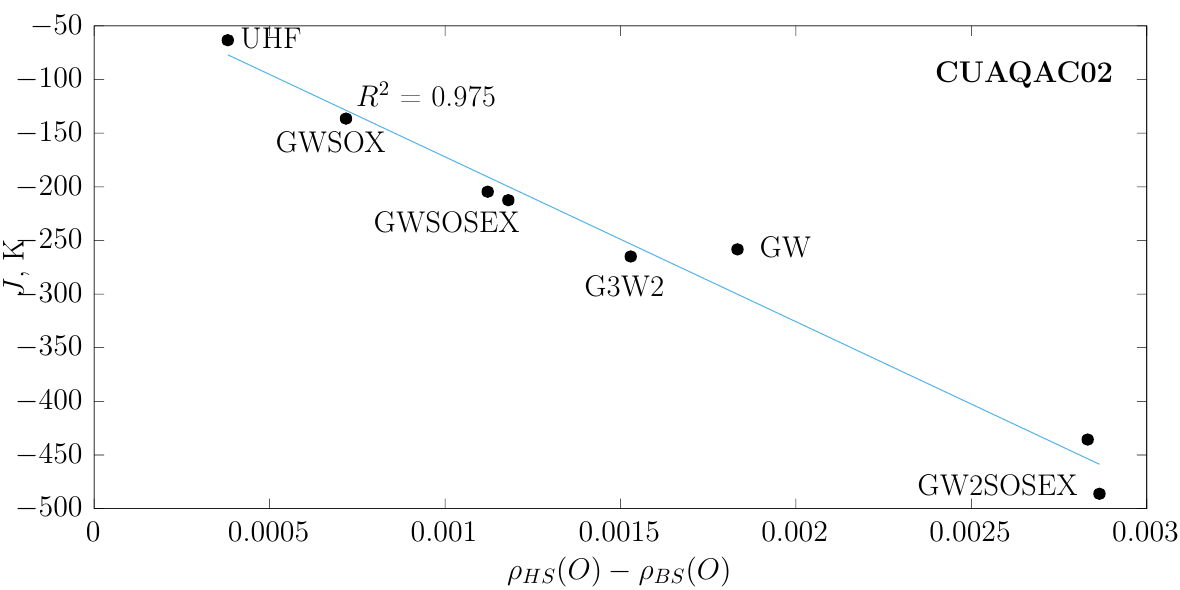}
  \includegraphics[width=9cm]{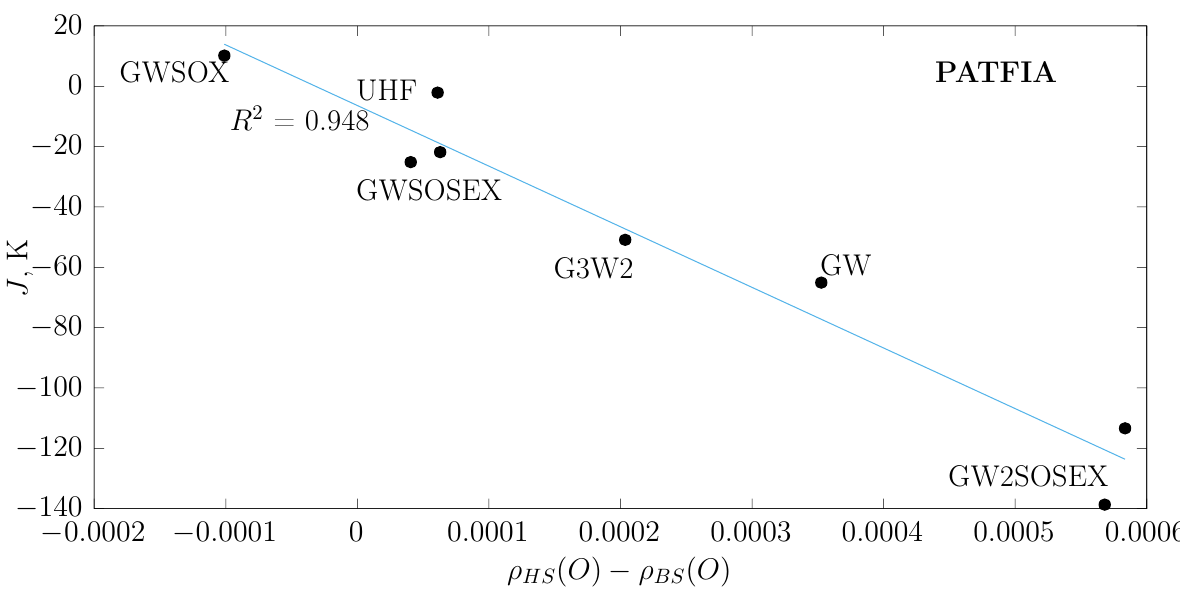}
\centering
\caption{Effective exchange couplings (vertical axis) plotted against the difference of the total density on oxygen atoms (using Lowdin's orthogonalized atomic orbitals) for molecules. 
         \protect\label{fig:ch_fit}
}
\end{figure}

As illustrated in Fig.~\ref{fig:ch_fit}, the differences between solutions in local electron densities on bridging oxygen atoms are 
almost perfectly correlated with observed changes in $J$ evaluated with different methods for all the molecules studied in this paper.  This also confirms that superexchange determines the overall strength of the effective exchange coupling. 
In particular, this analysis explains why GWSOX predicts the ferromagnetic coupling in PATFIA and relates it with the negative local charge difference on oxygen between HS and BS solutions. 

The effective exchange couplings evaluated with BS-scG3W2stat are almost identical to the scGW ones, 
which indicates that violation of crossing symmetry by GW does not play a significant role for magnetic interactions. 
The occupancies of SA-NOs (Fig.~\ref{fig:fe_NO1}) evaluated by scG3W2stat are the closest to scGW among all the methods shown. 
The role of crossing symmetry is additionally confirmed by making a comparison with BS-CCSD.  
CCSD contains bubble diagrams\cite{Scuseria:RPA:CC:2008}. 
The exponential treatment of the single excitations in CCSD is an efficient treatment of orbital relaxation, which is captured in the Green's function methods through full self-consistency.

BS-scGW J coupling estimates are very similar to the BS-CCSD ones, revealing the nature of the captured correlation and confirming that ladder diagrams do not play a significant role in the diagrammatic expansion.
EOM-SF-CCSD estimates are close to the BS-CCSD ones indicating that the ionic configurations described by EOM-SF-CCSD
in a balanced manner represent a weak or intermediate correlation. 
The weights of these configurations are small. Thus, not surprisingly, they can be perturbatively captured by the infinite diagrammatic series in BS-CCSD.  

J-coupling constants obtained from BS-DFT have been explored extensively in the previous decades. Such estimates are known to be sensitive to the employed functional, the percentage of HF exchange, and the range-separation parameter $\omega$.~\cite{Peralta:BS-DFT_benchmark:2011} 
The percentage of HF exchange and the range-separation parameter are usually strongly system-dependent.~\cite{Peralta:BS-DFT_benchmark:2011} 
This severe dependence on parameters is present in spin-flip time-dependent density functional theory (SF-TDDFT) as well with varying recommendations concerning best functionals and parameters.~\cite{Valero:SFDFT:11,Orms:magnets:17,Kotaru:Fe:SF-DFT:2023} 

We performed a simple test of BS-DFT with a few common functionals in comparison with the broken-symmetry Green's function and broken-symmetry wave-function approaches that we show in Table~\ref{tbl:J_sc_mol}. We confirm a strong dependence on functional types and their parametrizations. Changing functional from PBE to PBE0 results in the  J-coupling constants changing from -707.8 to -98.2 for PATFIA. Similarly, for CUAQAC02,  J-coupling from PBE0 is around 3 times smaller than from PBE, and for [Fe$_2$OCl$_6$]$^{2-}$ it is almost 2 times smaller. B3LYP performs reasonably for [Fe$_2$OCl$_6$]$^{2-}$ and PATFIA, but fails for CUAQAC02. Unlike for BS-DFT, trends observed for broken-symmetry wave-function and broken-symmetry Green's function methods are clear and systematic.

A direct comparison with the values extracted from experiments requires caution. 
Most of the values listed in Table~\ref{tbl:J_sc_mol} are obtained from the fit of 
temperature dependence of magnetic susceptibility to the Heisenberg model. 
While the Heisenberg model describes the case of $S=1/2$ for copper molecules exactly (the number of parameters equals the number of energy gaps), it may not be exact for higher spins, 
for which additional terms, such as biquadratic exchange, may be needed. 
In Ref. \cite{Molins:Fe2OCl6:magnet:2003} this error was quantified for [Fe$_2$OCl$_6$]$^{2-}$. For this compound, the biquadratic exchange strength is much smaller than the errors of the methods that we investigate in our work. 
Moreover, the experimental samples are often contaminated with impurities and free metal ions, 
which can affect the measured susceptibility.~\footnote{
For example, for the copper oxalate, the differences between samples as well as in their treatment 
were so large that only a rigorous 
theoretical study could find the less affected ones and 
derive the magnetic model through the Bloch formalism\cite{Pokhilko:spinchain}. }
Various ways of accounting for such impurities can give somewhat different experimental estimates. 
The original experimental paper\cite{lopez:PATFIA:05} honestly reports multiple issues 
with the fitting of the magnetic susceptibility as well as with the empirical derivation of the magnetic model. 
Due to these uncertainties, we focus only on a comparison between theories for PATFIA.

A comparison with other theories also requires caution with regard to the used geometry and the basis set. 
Dunning's cc-pVDZ basis set is already good enough for effective exchange couplings. 
Employing larger bases from the cc-pVXZ series gives only very marginal improvement of the order of very few K\cite{Orms:magnets:17,Pavel:OSFNO:2019,Pokhilko:spinchain}. 
The good performance of cc-pVDZ is likely due to the fact that the considered electronic states have 
nearly the same orbital occupancies. 
This is not the case of 6-31G*, which gives much larger errors.\cite{Truhlar:J:SFTDDFT:2011} 

The unsatisfactory performance of CASSCF for the effective exchange couplings is known well and is clear from the listed values in Table~\ref{tbl:J_sc_mol}. 
Inclusion of dynamic correlation by means of multireference CI, difference-dedicated CI, and selective CI 
improves the estimates. 
These results are fully consistent with our observations which 
explain the physical origin of the dynamic correlation in such compounds.

\subsection{Effective exchange couplings in solids and Neel temperatures}\label{sec:results_periodic_j}
\begin{table*} [tbh!]
  \caption{Effective exchange couplings (K) in solids evaluated with different methods. 
}
\protect\label{tbl:J_sc_solids}
\begin{tabular}{l|cc|cc}
\hline
\hline
NiO   & \multicolumn{2}{c|}{$3\times 3\times 3$} &  \multicolumn{2}{c}{$4 \times 4 \times 4$}  \\
\hline
                   & $J_1$   &  $J_2$  & $J_1$ & $J_2$    \\ 
BS-UHF             &   8.6   & -60.3   & 9.0  & -58.0    \\ 
BS-scGW            &   15.7  & -157.5  & 19.6 & -155.2    \\ 
BS-scGWSOX         &   12.2  & -104.4  & 16.0$^*$ & -102.1$^*$    \\ 
BS-scGWSOSEXstat   &   14.3  & -136.6  & 18.2$^*$ & -134.3$^*$     \\ 
BS-scGWSOSEXdyn    &   13.9  & -134.3  & 17.7$^*$ & -132.0$^*$     \\ 
BS-scGW2SOSEXstat  &   18.4  & -193.9  & 22.3$^*$ & -191.5$^*$     \\ 
BS-scG3W2stat      &   15.8  & -156.5  & 19.7$^*$ & -154.2$^*$    \\ 
\hline
MnO   & \multicolumn{2}{c|}{$3\times 3\times 3$} &  \multicolumn{2}{c}{$4 \times 4 \times 4$}  \\
\hline
BS-UHF             & -2.5  & -3.1  & -2.5 & -2.2  \\ 
BS-scGW            & -6.7  & -8.6  & -7.1 & -6.6  \\ 
BS-scGWSOX         & -4.6  & -5.9  & -5.0$^*$ & -3.9$^*$    \\ 
BS-scGWSOSEXstat   & -6.1  & -7.5  & -6.4$^*$ & -5.5$^*$     \\ 
BS-scGWSOSEXdyn    & -5.8  & -7.5  & -6.2$^*$ & -5.6$^*$     \\ 
BS-scGW2SOSEXstat  & -7.9  & -9.5  & -8.2$^*$ & -7.5$^*$     \\ 
BS-scG3W2stat      & -6.8  & -8.4  & -7.1$^*$ & -6.4$^*$    \\ 
\hline
\hline
\end{tabular}

$^*$ Estimate from Eq.\ref{eq:J_est}.
\end{table*}

\begin{table*} [tbh!]
  \caption{Previously published effective exchange couplings (K) in solids evaluated with DFT. 
}
\protect\label{tbl:J_sc_solids_DFT}
\begin{tabular}{l|cc|l|cc}
\hline
\hline
 NiO  & $J_1$  & $J_2$ & MnO & $J_1$ & $J_2$ \\
\hline
LDA$^a$   & 138   &  -827     &        &        &    \\ 
B3LYP$^a$ & 28    &  -310     & B3LYP$^c$  & -9.8  &  -20.4       \\ 
PBE$^b$   & 14    &  -516     & PBE$^b$    & -17.6 &  -27.7   \\ 
PSIC$^b$  & 38    &  -287     & PSIC$^b$   & -9.3  &  -14.1   \\ 
ASIC$^b$  & 60    &  -522     & ASIC$^b$   & -14.9  &  -21.0   \\ 
PBE0$^b$  & -72   &  -86      &        &        &    \\ 
\hline
\hline
\end{tabular}

$^a$ Ref. \cite{Martin:NiO:exchange:2002}

$^b$ Ref. \cite{Majumdar:NiO:MnO:DFT:J:2011}

$^c$ Ref. \cite{Feng:MnO:CoO:b3lyp:2004}
\end{table*}

The effective exchange couplings\cite{note:prev_J_solid} evaluated with different methods for 
NiO and MnO are shown in Table~\ref{tbl:J_sc_solids}. 
We showed that effective exchange couplings, unlike band gaps, 
converge very fast to the thermodynamic limit\cite{Pokhilko:BS-GW:solids:2022}. 
For NiO and MnO, the $4\times 4 \times 4$ k-point grid already gives a fully converged estimate of $J$. 
Since THC-SOX has a cubic cost with respect to the number of k-points, 
going to $4\times 4 \times 4$ k-point grid is computationally expensive. 
$J$ is a local property and since SOX decays fast with a distance in the real space, 
we estimate the values of $J$ at the thermodynamic limit as
\begin{gather}
J(4\times 4\times 4) = J(3\times 3\times 3) + J_{GW}(4\times 4\times 4) \nonumber \\
- J_{GW}(3\times 3\times 3)
\protect\label{eq:J_est}
\end{gather}

The obtained estimates of $|J|$ in NiO and MnO follow the same trends as in molecules. 
The inclusion of screening increases the magnitude of both ferromagnetic and antiferromagnetic interactions from UHF to scGW. 
The bare SOX term reduces $|J|$. Replacing one bare interaction with a screened interaction $W$ (scGWSOSEX) 
leads to $|J|$ estimates that are between scGW and scGWSOX. 
scGW2SOSEX gives larger values of $|J|$ than scGW. 
Static and dynamical treatment of $W$ in the SOX term gives very close estimates of $J$, 
which again validates the quality of the static approximation for solids. 
The scG3W2stat estimates of $J$ are very close to scGW, 
which also indicates that the low-order insertions of the screened vertex functions into self-energy 
do not change the estimates of effective exchange couplings. 

\begin{figure}[!h]
  \includegraphics[width=9cm]{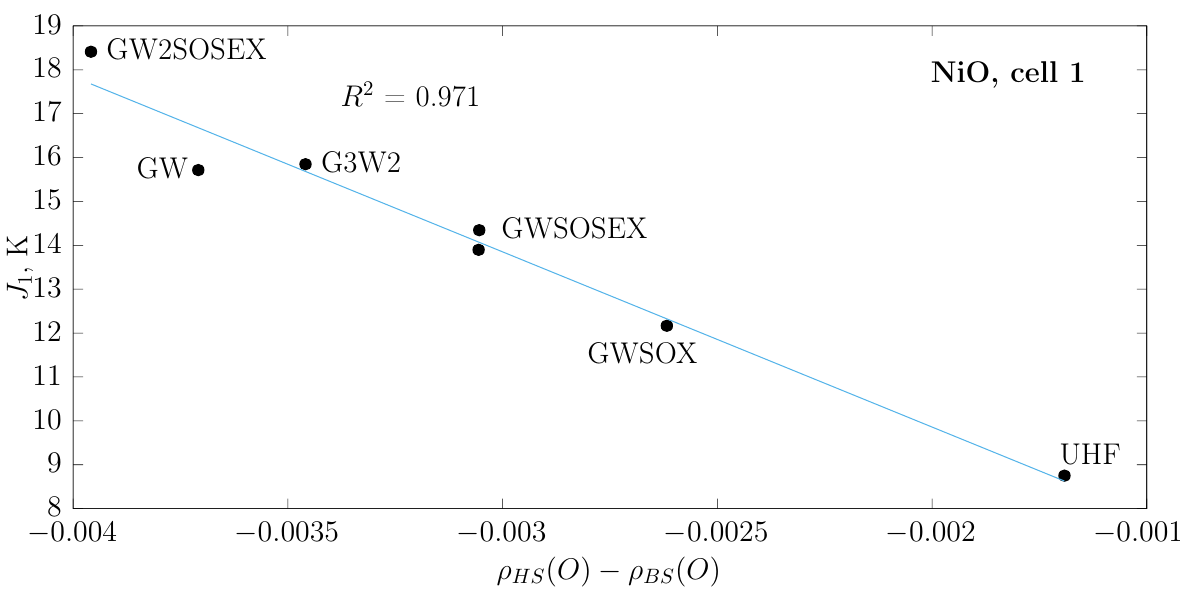}
  \includegraphics[width=9cm]{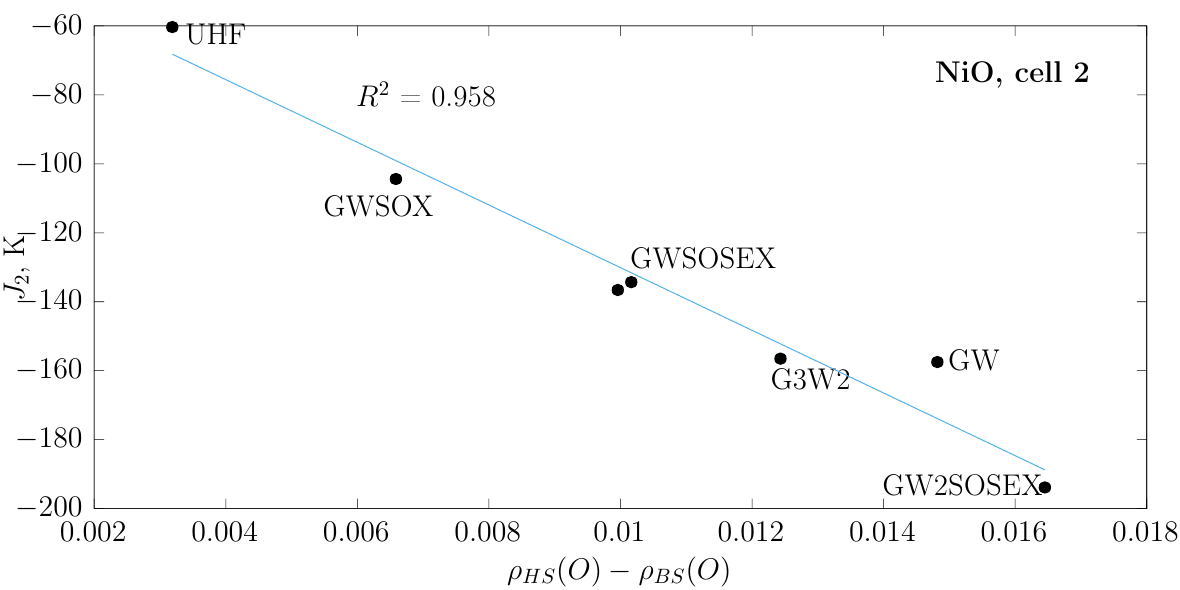}
\centering
\caption{Effective exchange couplings (vertical axis) evaluated with $3\times 3\times 3$ k-grid plotted against the difference of the total density on oxygen atoms (using Lowdin's orthogonalized atomic orbitals) for NiO. 
         \protect\label{fig:NiO_ch}
}
\end{figure}
\begin{figure}[!h]
  \includegraphics[width=9cm]{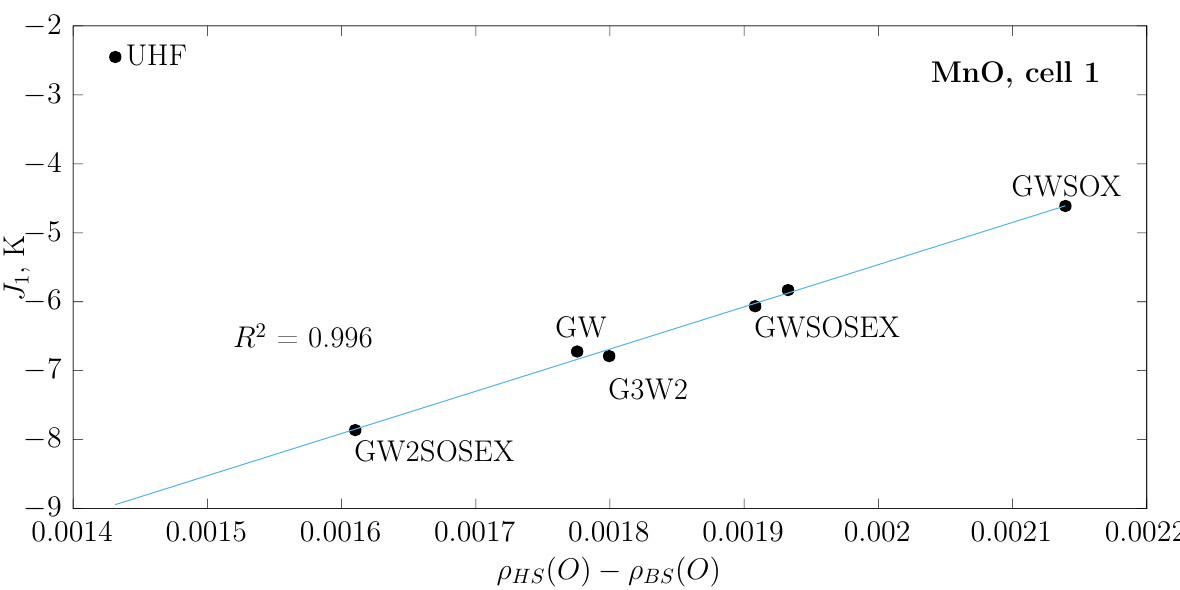}
  \includegraphics[width=9cm]{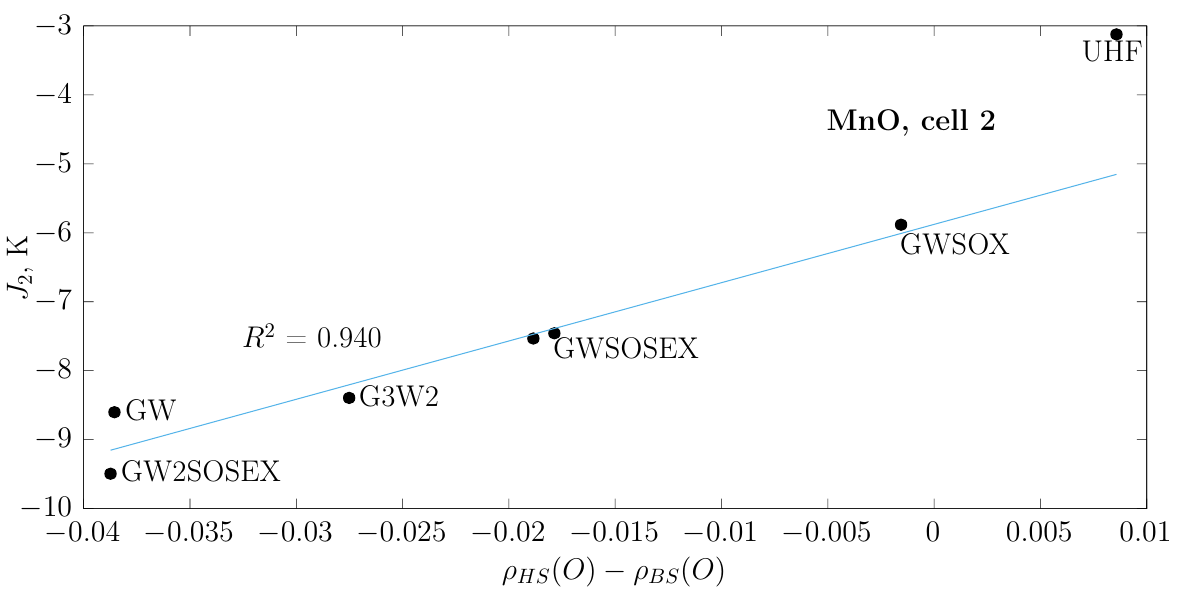}
\centering
\caption{Effective exchange couplings (vertical axis) evaluated with $3\times 3\times 3$ k-grid plotted against the difference of the total density on oxygen atoms (using Lowdin's orthogonalized atomic orbitals) for MnO. The point corresponding to UHF was not taken into the linear fit. 
         \protect\label{fig:MnO_ch}
}
\end{figure}
Since the structure of natural orbitals of these solutions strongly depends on a particular k-vector\cite{Pokhilko:NO:correlators:2023}, 
we considered instead changes of the local real-space Lowdin charges on the oxygen atom in the starting unit cell, 
which is consistent with the measure we used for molecules in the previous section. 

In NiO, consistently with molecules, magnitudes of both ferromagnetic and antiferromagnetic couplings increase 
as the magnitude of the difference in charges increases (Fig.\ref{fig:NiO_ch}). 
In cell 2, the difference in electron density on oxygen atoms between the HS and the BS solutions is positive, 
which is consistent with molecular antiferromagnets and which corresponds to formations of partial multi-center multielectron bonds. 
In cell 1, the difference in electron density on oxygen atoms between the HS and the BS solutions is negative. 
The quality of the linear fit indicates that superexchange is the cause of both 
ferromagnetic and antiferromagnetic interactions, 
which are expected from semiempirical Goodenough--Kanamori rules\cite{Goodenough:direct_exchange:1960,Kanamori:exchange_mechanisms:1959,GK:rules:summary}, 
assumptions of which we explored in our previous work\cite{Pokhilko:NO:correlators:2023}.  
The sign of the slope of the linear fit is the same as for molecules for both cells.

In MnO, the trends are qualitatively different from NiO and molecules, which could indicate a different physics. 
First, UHF points in Fig.\ref{fig:MnO_ch} are outliers from the expected dependence; 
a linear fit with good quality is possible only for the correlated data points. 
The deviation of UHF from the correlated trend is especially significant for cell 1, 
describing $J_1$ nearest-neighbor interaction between Mn$^{2+}$, 
which is the case of different competitive exchange mechanisms\cite{Kanamori:exchange_mechanisms:1959}.  
This change of character is also consistent with different signs of differences in 
local spin correlators on oxygen for UHF and GW in cell 1 that we observed previously\cite{Pokhilko:BS-GW:solids:2022}. 
The magnitude of $|J_1|$ in MnO decreases with an increase in the charge transfer, 
which is the opposite of what one would expect from a usual superexchange mechanism.  
The signs of the density difference on oxygen are different for UHF and the correlated methods in cell 2, 
which also points to a change in physics with correlated methods. 
The sign of the linear slope is the opposite from NiO and molecules; 
the signs of the density difference between HS and BS solutions in different cells are also the opposite from NiO cases. 

Table~\ref{tbl:J_sc_solids_DFT} shows the previously published DFT estimates for solids, from which a severe dependence of $J$ on a chosen DFT functional is clear. Not only magnitudes of $|J|$, but also signs of interactions are functional dependent.  For instance, going from PBE to PBE0 reverts the sign of the nearest-neighbor interactions. Due to a lack of systematic improvement and huge variations of estimates, Neel temperatures have not been successfully evaluated before with DFT functionals.

\begin{table*} [tbh!]
  \caption{Ratios of effective exchange couplings, Neel temperature estimates (units of $J_2$ and K) for NiO and MnO evaluated with different methods. An extrapolation to the infinite order of the ratio estimate ($q_\infty$) is estimated from the linear fit of the last 5 orders, $q_{6}$--$q_{10}$.
}
\protect\label{tbl:TN_sc_solids}
\begin{tabular}{l|ccc}
\hline
\hline
NiO                & $J_1/J_2$   &  $g_{10}(J_1/J_2, 1)$  & $g_{10}(J_1, J_2)$, K    \\ 
\hline
BS-UHF             & -0.156  & 1.57 & 91 \\
BS-scGW            & -0.126  & 2.00 & 310 \\
BS-scGWSOX         & -0.157  & 1.62 & 166 \\
BS-scGWSOSEXstat   & -0.135  & 1.89 & 253 \\
BS-scGWSOSEXdyn    & -0.134  & 1.90 & 251 \\
BS-scGW2SOSEXstat  & -0.116  & 2.08 & 399 \\
BS-scG3W2stat      & -0.128   & 1.98  & 306 \\
\hline
MnO                & $J_1/J_2$   &  $q_{10}(J_1/J_2,1)$  & $q_{\infty}(J_1, J_2)$, K    \\ 
\hline
BS-UHF             & 1.160  & 50.32  & 107 \\
BS-scGW            & 1.064  & 47.52 & 305 \\
BS-scGWSOX         & 1.264  & 53.36 & 202 \\
BS-scGWSOSEXstat   & 1.167  & 50.52 & 268 \\
BS-scGWSOSEXdyn    & 1.109  & 48.82 & 263 \\
BS-scGW2SOSEXstat  & 1.090  & 48.26 & 351 \\
BS-scG3W2stat      & 1.109  & 48.83 & 303 \\
\hline
\hline
\end{tabular}
\end{table*}
\begin{table} [tbh!]
  \caption{Experimental estimates of Neel temperature (K).   
\protect\label{tbl:Neel_T}}
\centering
\begin{tabular}{c|c}
\hline
\hline
      & Exp \\
NiO   & 530$^a$  516$^b$  524.5$^c$  523$^d$ 520$^g$ \\ 
MnO   & 118$^f$  122$^g$ 116$^g$ \\
\hline
\hline
\end{tabular}

$^a$: Reference\citenum{Lindgard:NiO:2009}

$^b$: Reference\citenum{Vernon:NiO:1970}

$^c$: Reference\citenum{Seehra:NiO:1984}

$^d$: Reference\citenum{Balagurov:NiO:MnO:2016}

$^f$: Reference\citenum{Goncharenko:MnO:FeO:neutron_dif:2005}

$^g$: Reference\citenum{Kubo:antiferromagnetism:1955}
\end{table}
The estimated Neel temperatures are shown in Table~\ref{tbl:TN_sc_solids}. 
Due to Eq.~\ref{eq:homogen_cn}, the estimated Neel temperatures depend not only on 
the magnitude of the $|J_1|$ or $|J_2|$, but also on their ratios. 
For NiO, the $J_1/J_2$ ratio evaluated with scGWSOX is similar to the one found from UHF. 
When screened interactions are included in the SOX term, the closer the ratios become to scGW. 
For the statically screened scG3W2stat, the effective exchange constants and their ratios 
are very close to the ones evaluated with scGW. 
The $J_1/J_2$ ratios found from scGW and scG3W2stat are between the ones found from scGWSOSEX and scGW2SOSEX, 
which is fully consistent with the trends for the magnitudes of $|J|$ and local charge differences. 

For MnO, the behavior of ratios is more complicated because of both more sophisticated trends in $|J|$ and 
bigger sensitivity of the ratios with respect to errors in $J$. 
Nonetheless, the resulting estimates of the Neel temperature follow the same trends as $J_2$ for both NiO and MnO. 
The resulting estimates of $T_N$ underestimate the experimental measurements for NiO and 
overestimate the experimental estimates for MnO.  
The underestimation of $T_N$ is consistent with the trends in molecular systems, 
where the fully self-consistent correlated Green's function methods underestimate $J$. 
DDCI3 usually recovers the dynamic correlation well and is sometimes considered as a reference standard for 
effective exchange couplings. 
In CUAQAC02, DDCI3 delivers estimates that are close to the experimental values. 
While it is not possible to compare $J$ in solids with the experiment directly, 
we can estimate the Neel temperatures from them, which we did in our previous work\cite{Pokhilko:Neel_T:2022}. 
$T_N$ from DDCI3 is closer to the experiment than scGW showing that the effective exchange couplings in NiO are likely underestimated with the Green's function methods that we investigated. 

The case of MnO is harder to rationalize.  
Previously, we hypothesized that there is a possible vibronic screening component due to a high likelihood of 
simultaneous structural phase transition\cite{Pokhilko:Neel_T:2022}.  
In this work, we also observe that the structure of electron correlation is more complicated than in other compounds.  
So this overestimation may stem from more complicated mechanisms of exchange interactions 
including higher-order electronic and vibronic effects.  

\section{Conclusions and future work}\label{sec:conclusions}

We have extended the THC algorithms for SOX with static and dynamic treatment of screening to both molecules and solids. 
We implemented these new algorithms in an efficient MPI-parallel manner. 
We showed that the reduction of scaling by THC has different regimes determined by $O(n_t N^2 n_{AO}^2 n_k^2)$ and $O(n_t N^2 n_{AO} n_k^3)$ contractions. 
The overall cost reduction by THC contractions allowed us to perform the largest scGWSOX, scGWSOSEX, scGW2SOSEX, scG3W2 calculations up to date for transition-metal complexes and periodic solids.
We have demonstrated that THC can be used reliably for energy differences in transition-metal compounds. 
We have also confirmed that these Green's function methods undergo symmetry breaking for such open-shell compounds. 
For both molecules and solids, we have successfully used the developed methods within the broken-symmetry approach to investigate the influence of screening present in the SOX term on the effective magnetic exchange couplings. 
We showed that the signed dynamical components of energies are ordered as  
$E_{dyn}(GWSOX) > E_{dyn}(GWSOSEXstat) > E_{dyn}(G3W2stat) > E_{dyn}(GW2SOSEXstat) > E_{dyn}(GW)$, 
which we explained through a dependence on an effective dielectric constant. 
The ordering of self-energies largely follows the ordering of the dynamical energies. 
Using the full dynamical $W$ and the statically screened interaction in the SOX term results in big changes in 
$E_{dyn}$ originating from both real and imaginary components of self-energy. 

We have confirmed that a non-self-consistent evaluation of the effective exchange couplings produces unphysical results. 
This is not surprising because non-self-consistent methods violate conservation of energy making the energy evaluation ambiguous. 
In contrast, fully self-consistent methods deliver physically meaningful results and uncover the origin of electron correlation in such compounds. 
In all of the considered molecules and solids the magnitude of $|J|$ increases from UHF to scGW. 
The physical origin of this change is in superexchange, which can be understood through a partial charge transfer 
between ligands and metals in the localized picture or, equivalently, 
through a formation of partial multi-electron multi-center bonds.  
This results in deviations of natural orbital occupations away from 1 for the antiferromagnetic states and 
broken-symmetry solutions. 
The frontier natural occupations for the ferromagnetic solutions are close to 1 and 
are protected by Pauli principle from the same-spin transfer of population.  
The incorporation of electronic screening effectively lowers the energies of the charge-transfer configurations, 
making their presence more prominent in the wave functions, Green's functions, and density matrices. 
Since energies are determined by one-particle and two-particle density matrices, 
this leads to an increase of $|J|$ from BS-UHF to BS-scGW.

The incorporation of SOX reduces $J$ when compared to BS-scGW. 
The magnitude of the change depends on the number of screened interactions included in the SOX term. 
The reduction is the biggest for BS-scGWSOX with both bare interaction lines, 
and it is smaller for BS-scGWSOSEX.  
The values of $J$ found from BS-G3W2stat are almost the same as the ones found from BS-scGW. 
The BS-scGW2SOSEX estimates of $|J|$ are larger than the ones from BS-scGW. 
We explain all the observed trends through the treatment of the effective dielectric constant in the SOX term 
consistently with the analysis of self-energies. 
We observe that unlike total energies, 
the effective exchange couplings found from SOX with static and dynamic screening are almost identical 
for all the systems, which indicates that static approximation is an excellent approximation for the 
total energy differences. 
We confirm our reasoning by analyzing natural orbitals and their occupations. 
We show that the values of the effective exchange couplings closely follow a linear dependence 
with respect to the local electron density differences on oxygen atoms for both molecules and solids. 
Such an excellent linear dependence shows that the origin of the effective exchange is mainly due to superexchange, 
which was suspected previously\cite{Hoffman:exchange:1975,Gudel:Fe2O:analyt,Goodenough:direct_exchange:1960,Kanamori:exchange_mechanisms:1959,GK:rules:summary}. 
The only deviations occur for MnO by the slope sign and by the quality of the fit (BS-UHF is the outlier) indicating 
that the origin of effective exchange in MnO is more complicated, 
which may occur due to both higher-order electronic and vibronic effects.  

An excellent agreement of BS-GW with BS-scG3W2stat and a good agreement between BS-scGW, BS-CCSD, EOM-SF-CCSD 
reveals the nature of the effective exchange couplings from a diagrammatic perspective. 
Neither an insertion of a single screened line into 
the GW self-energy nor CCSD ladders change the BS-scGW estimate noticeably. 
This means that the effective couplings are determined by the GW diagrams rather 
than the higher-order diagrams generated by vertex insertions into self-energy. 
This means that the further improvement lies in the improvement of the polarization propagator, 
on which we will focus in the nearest future. 
This diagrammatic analysis limits a potential applicability of the finite-order methods  
due to the lack of interaction renormalization 
necessary for a correct description of electronic screening. 

Finally, we applied the high-temperature expansion to the derived Heisenberg Hamiltonian for NiO and MnO and used the obtained expansion coefficients to estimate the Neel temperatures. 
Although the Neel temperatures depend on both the ratio and magnitude of the effective exchange couplings, 
we found that the Neel temperatures follow the trends present for the effective exchange couplings.  
The estimated Neel temperatures in NiO are smaller than the experimental ones, 
which is consistent with the molecular trend of underestimation of the magnitude of $|J|$.  
The estimated Neel temperature in MnO exceeds the experimental one, 
which again reveals the complexity of magnetic interactions in this compound.

\section*{Acknowledgments}
P. P. and D. Z. acknowledge support from the National Science Foundation under grant number CHE-2154672.

The Flatiron Institute is a division of the Simons Foundation.

\section*{Authors Declaration}
The authors have no conflicts to disclose.

\section*{Supplementary Material}
Weights of the 2p oxygen's AO orbitals in SA-NOs; convergence of the high-temperature expansion coefficients with respect to the order of expansion.

\section*{Data Availability}
The data supporting the findings of this study are available within the article and its supplementary material.

\renewcommand{\baselinestretch}{1.5}


\begin{thebibliography}{100}
\bibitem{note:H_def}
There are different notations with different ranges of summations and with
  different prefactors and signs. Such details are important for comparisons
  between different papers and must be declared.
\bibitem{note:prev_J_solid}
The J estimates for NiO and MnO are slightly different from our previous
  estimates. We investigated the origin of this small discrepancy and found
  that it is due to an aggressive integral prescreening of RI integrals for
  solids computed by PySCF. The current paper does not use these prescreened
  integrals in the final tabulated estimates.
\end{thebibliography}
\end{document}